\documentclass[superscriptaddress,aps,twocolumn,showpacs,amsmath,amssymb]{revtex4-1}
\pdfoutput=1

\usepackage{graphicx}% Include figure files
\usepackage[outdir=./]{epstopdf}
\usepackage{dcolumn}% Align table columns on decimal point
\usepackage{bm}% bold math
\usepackage[version=4]{mhchem}
\usepackage{comment}
\usepackage{tikz}
\usepackage{varwidth}
\usepackage{ragged2e}
\usepackage[english]{babel}

\newcommand {\apgt} {\ {\raise-.5ex\hbox{$\buildrel>\over\sim$}}\ }
\newcommand {\aplt} {\ {\raise-.5ex\hbox{$\buildrel<\over\sim$}}\ }

\begin{document}

\title{Loading a linear Paul trap to saturation from a magneto-optical trap} 

\author{J.~E.~Wells}
\affiliation{W. M. Keck Science Department of Claremont McKenna, Pitzer, and Scripps Colleges, Claremont, California 91711}
\affiliation{Department of Physics, University of Connecticut, Storrs, Connecticut 06269}
\author{R.~Bl\"{u}mel}
\affiliation{Department of Physics, Wesleyan University, Middletown, Connecticut 06459}
\author{J.~M.~Kwolek}
\affiliation{Department of Physics, University of Connecticut, Storrs, Connecticut 06269}
\author{D.~S.~Goodman}
\affiliation{Department of Physics, University of Connecticut, Storrs, Connecticut 06269}
\affiliation{Department of Sciences, Wentworth Institute of Technology, 
Boston, Massachusetts 02115}
\author{W.~W.~Smith}
\affiliation{Department of Physics, University of Connecticut, Storrs, Connecticut 06269}
  
\date{\today}

\begin{abstract}
We present experimental measurements of the
steady-state ion number in a linear Paul trap (LPT)
as a function of the ion-loading rate. These measurements,
taken with (a) constant Paul trap stability parameter $q$,
(b) constant radio-frequency (rf) amplitude, or
(c) constant rf frequency,
show nonlinear behavior. At the loading rates achieved in this
experiment, a plot of the steady-state ion number as a function
of loading rate has two regions: a monotonic rise (region I)
followed by a plateau (region II). Also described are simulations
and analytical theory which match the experimental results.
Region I is caused by rf heating and is fundamentally
due to the time dependence of the rf Paul-trap forces.
We show that
the time-independent pseudopotential, frequently used
in the analytical investigation of trapping experiments,
cannot explain region I, but
explains the plateau in region II and can be used to
predict the steady-state ion number in that region.
An important feature of our experimental LPT is the existence
of a radial cut-off $\hat R_{\rm cut}$ that
limits the ion capacity of our LPT and features
prominently in the analytical and numerical analysis
of our LPT-loading results. We explain the dynamical
origin of $\hat R_{\rm cut}$ and relate it to the
chaos border of the fractal of non-escaping trajectories
in our LPT. We also present
an improved model of LPT ion-loading as a function of time.
\end{abstract}

     % Plasma heating by rf fields 


\maketitle

\section{Introduction} 
\label{INTRO} 

The loading dynamics of the linear Paul trap (LPT) are 
of interest in relation to recent work measuring the total 
charge exchange and elastic collision rate of 
atomic ions with their parent atoms in a 
hybrid trap \cite{Lee:2013,Goodman:2015}. In these hybrid-trap
measurements of collisions between atoms and 
dark ions, those without optically accessible 
transitions, the fluorescence of the atoms in 
the MOT was monitored in the presence of trapped 
ions. Knowledge of both the number of trapped ions and 
the size of the ion cloud is necessary for finding 
the total interaction rate. Both of these papers 
attempted to create a model for the loading of ions 
into a Paul trap. However, neither group was able 
to create a satisfactory model of Paul trap loading. 
This paper aims to understand and model the loading 
of a linear Paul trap at the conditions used in current 
hybrid trap experiments.

Hybrid traps consist of a neutral atom trap coincident 
with an ion trap. A variety of neutral atom traps have been 
used in hybrid trap experiments: magnetic traps \cite{Zipkes:2010a,Zipkes:2010b,Schmid:2010}, 
magneto-optical traps (MOT) \cite{Grier:2009,Rellergert:2011,Hall:2011,Ravi:2012,Deiglmayr:2012,Sivarajah:2012,Hall:2013,Lee:2013,Haze:2013,Smith:2014,Haze:2015,Goodman:2015}, and optical dipole traps \cite{Harter:2012,Krukow:2016,Meir:2016}. In contrast, every hybrid trap experiment listed above used a Paul trap to hold the ionic species, except \cite{Deiglmayr:2012}, which used an octupole trap.

In hybrid trap experiments for dark ions and their parent atoms, the ion-atom 
total elastic and inelastic interaction rate per atom $\gamma_\mathrm{ia}$ is given by \cite{Goodman:2015}
\begin{equation}
\gamma_\mathrm{ia}=\frac{k_\mathrm{ia}N_IC}{V_\mathrm{ia}}.
\label{eq:rate}
\end{equation}
In this equation $k_\mathrm{ia}$ is the total elastic and charge-exchange collision rate constant, 
$N_I$ is the average number of trapped ions overlapping the atom cloud, $C$ is a function that describes the 
concentricity of the atom and ion clouds, and $V_\mathrm{ia}$ is the effective 
overlap volume of the two clouds \cite{Goodman:2015}. 

In these experiments the ion trap is saturated, 
which has two benefits. The first is that this
maximizes the rate and thus the experimental resolution of the 
rate constant, $k_\mathrm{ia}$, because the ion number
and volume will be as large as possible.    

The second benefit to a saturated ion trap is that the ion number, the
concentricity, and the overlap volume will all be time independent on average.  In the case of dark ions, the ion number
can only be measured destructively; this measurement would be extremely difficult if 
the ion number were time dependent. For bright ions, those with optical transitions, 
these quantities could be measured in a time dependent 
way using the fluorescence. Typically, however, bright ion
experiments have focused on the charge-exchange rate constant and have not measured 
the total rate. If the technique described in \cite{Lee:2013,Goodman:2015} were 
used in conjunction with the methods
for measuring the charge exchange rate
constant, then the elastic rate constant could
be found as well.   

Neither of the dark ion experiments developed a satisfactory model
 for predicting the steady-state population of the saturated Paul trap 
 as a function of the ionization rate. In the model used in Lee, \emph{et \ al.} \cite{Lee:2013},
an \emph{ad hoc} term was required to prevent the ion number from becoming infinite
at large photoionization intensities and $\gamma_\mathrm{ia}$ was proportional 
to the Paul trap loss rate $\ell$. In Goodman, \emph{et \ al.} \cite{Goodman:2015}, 
we found experimentally that $\gamma_\mathrm{ia}$ was not proportional to $\ell$. We also 
found, by considering how the number of atoms in the MOT
depended on the photoionization intensity, that the equation for the ion number 
as a function of photoionization intensity is naturally finite
as the photoionization intensity tends to infinity.   

\begin{figure}
\centering
\includegraphics[scale=0.5,angle=0]{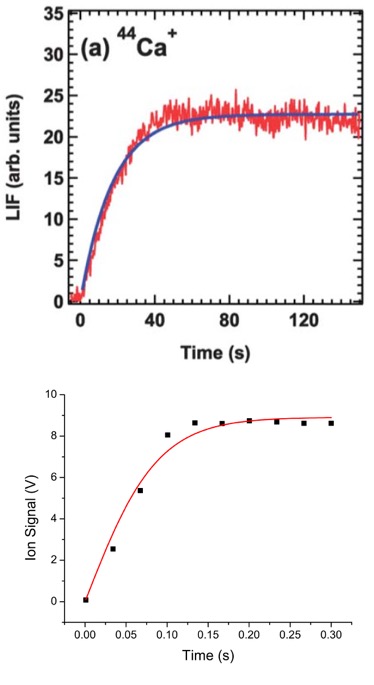}
\caption{\label{fig:tradfit} (Color online) 
Fits of ion loading as a function of time based on (\ref{eq:tradmod}) overshoot the rise, undershoot the knee, and overshoot the plateau. This is the case regardless of whether the ion number is measured via fluorescence at low loading rates (top, reproduced from Ref.~\cite{Kitaoka:2013} with permission from The Royal Society of Chemistry) or measured using a channel electron multiplier at high loading rates from a hybrid trap by our group (bottom). Where the error bars (showing statistical errors) are not seen, they are smaller than the corresponding plot symbols. }
\end{figure}

Ion loading in a Paul trap as a function of time is typically fit to the solution of
\begin{equation}
\label{eq:tradmod}
\frac{dN(\tau)}{d\tau}=\Lambda-\ell_1N(\tau)-\ell_2N(\tau)^2,
\end{equation}
where $N$ is the number of trapped ions, $\tau$ is the time, $\Lambda$ is the loading rate,
$\ell_1$ is the one-body loss rate, and $\ell_2$ is the two-body loss rate. All
quantities in (\ref{eq:tradmod}) are in SI units. The model (\ref{eq:tradmod}) is used, e.g., in
\cite{Lee:2013,Goodman:2015,Kitaoka:2013,Hasegawa:2015},
though sometimes $\ell_2$ is set to zero without significantly affecting the fit. The loading rate
$\Lambda$ is constant and set by the ion source. The one-body and 
two-body loss rates $\ell_1$ and $\ell_2$ are usually assumed to be independent 
of the loading rate. However, in Goodman, \emph{et\ al.} \cite{Goodman:2015}
we found that not to be the case. Additionally, while this model can fit well, 
as in \cite{Lee:2013,Goodman:2015}, in
other cases it overshoots the rise, undershoots the knee of the curve, and overshoots
the plateau, as can be seen in Fig.~\ref{fig:tradfit}. This trend is present even in experiments that
do not have a hybrid trap and therefore do not load from a MOT (see, e.g., Ref.~\cite{Kitaoka:2013}).

Neglecting the two-body loss mechanism in (\ref{eq:tradmod}),
i.e., for $\ell_2=0$, the traditional model from the solution of (\ref{eq:tradmod}) predicts that 
the steady-state ion number is directly proportional to the loading rate.
In our previous work, Bl{\"u}mel, \emph{et\ al.} \cite{Blumel:2015}, simulations 
showed that the steady-state ion number depends on the loading rate 
non-monotonically. There are four regions in a plot of the steady-state ion
number as a function of loading rate. At small loading rates, the steady-state ion
number increases monotonically, but nonlinearly, in region I. In region II, the steady-state ion number 
plateaus. Region III is characterized by a dip, where the steady-state ion number
decreases with increasing loading rate. Finally, in region IV, the steady-state ion number once again
increases monotonically with the loading rate, but much faster than in region I, and with a power that
does not agree with the predictions of the model defined in (\ref{eq:tradmod}). This
behavior was present for simulations of both the linear and three-dimensional Paul trap geometries.

Despite being in use for over 50 years \cite{Paul:1990} in many disciplines, 
including mass spectrometry, biology, chemistry, and physics, the fundamental 
loading behavior of the Paul trap is still unknown. This work examines the loading behavior 
of the Paul trap experimentally, analytically, and with simulations in regions I and II, where 
the loading rates are achievable experimentally in every hybrid trap system and many 
mass spectrometry systems. 

This paper is 
organized as follows: Section II describes the experimental apparatus and technique. Section 
III describes the experimental results and Sections IV and V compare those results with 
the simulations and the analytical models. We conclude in Section VI.

\section{Apparatus and Method}
\label{APP}

\begin{figure}
\centering
\includegraphics[width=\columnwidth]{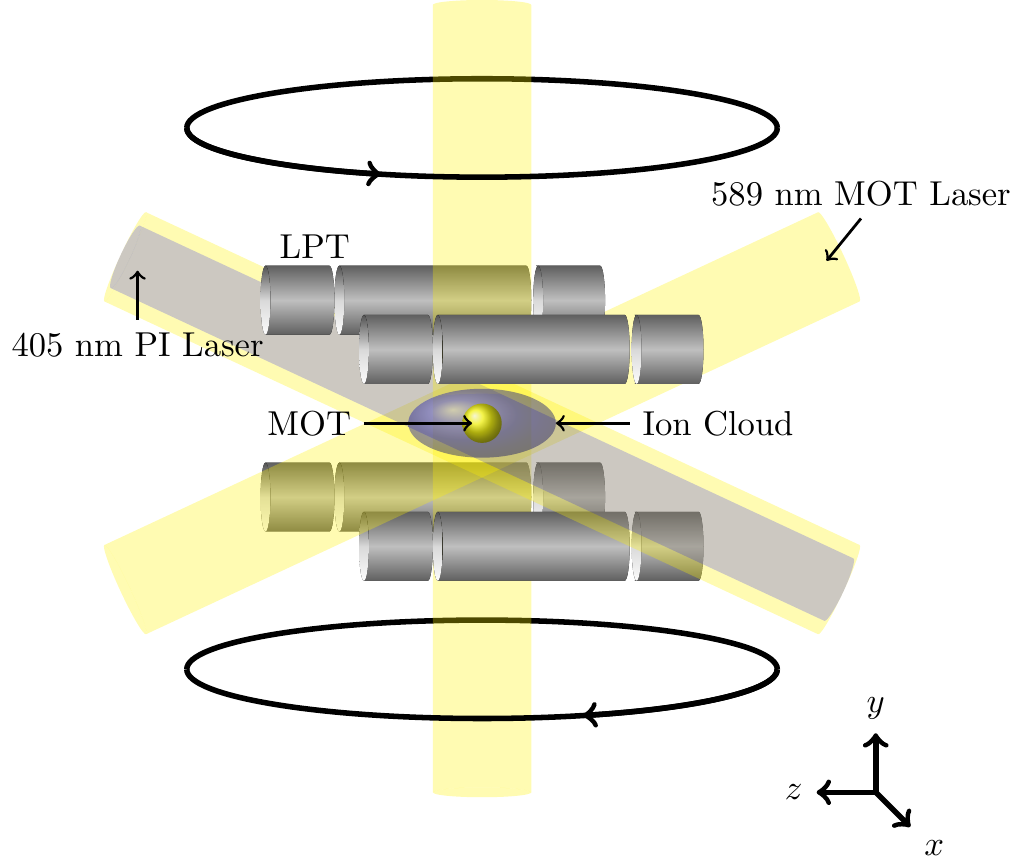}
\caption{\label{fig:app} (Color online) 
A diagram of the hybrid apparatus used in these experiments. The anti-Helmholtz coils, with current directions shown by arrows, are located outside of the vacuum chamber (not shown).}
\end{figure}

The apparatus used in this experiment has been described elsewhere \cite{Goodman:2015}, 
so we will only briefly describe it here. A diagram of our system can be seen in Fig.~\ref{fig:app}.
Our hybrid trap uses a linear radio frequency (r.f.) Paul trap with a segmented 
design as the ion trap and a sodium MOT as the neutral trap. The 
MOT is a vapor cell design where the background sodium vapor is
generated by a getter source. Even with constant 
loading from the getters, the background pressure is 
below $\approx 0.1 \times 10^{-9}\,$Torr.  There are two MOT transitions
in sodium involving different hyperfine levels \cite{Raab:1987}. We use the type-II MOT transition 
to take advantage of the higher trapped atom 
number achieved using that transition. The MOT is 
made by retro-reflecting the three trapping beams
and the repumper is obtained from a sideband put on the 589-nm 
beam using an electro-optic modulator prior to 
splitting them into three beams. The anti-Helmholtz coils
are located outside of the vacuum chamber. 

The MOT can be characterized in two ways: 
using a photomultiplier tube (PMT) 
or using a CMOS camera. Both the PMT and the CMOS 
camera can be used to measure the trapped atom number, but the camera can also 
be used to measure the MOT radius. The MOT has a 1/$e$-density radius 
of $\approx 0.75$ mm and holds on the order of $10^7$ atoms. 
The PMT yields temporal information 
about the loading of the MOT. 
 
The ions are created by a two-step process: atoms are resonantly
excited to the 3P state by the trapping laser
beams of the MOT, then they are ionized by a second laser beam at 
$405\,$nm. The first resonant step assures that the sample of loaded ions is pure,
 from the same species as the MOT neutrals.The beam size is fixed 
 at a $1/e$-intensity radius of $1.8\,$mm,
so the beam is always larger than the MOT and the intensity is 
approximately uniform over the size of the MOT. The photoionization
intensity is controlled by changing the power of the $405\,$nm beam.

The LPT consists of four cylindrical segmented rods, 
each having three segments, that lie along the long edges of a 
square prism. The short segments at either end of the rods are known 
as the endcaps and provide axial confinement when DC voltages are applied to them. The length of these 
segments is $17.0$ mm. The longer central segment of the rod provides 
the radial confinement using r.f. voltages applied to the diagonal pairs of rods. The length of this segment is $2z_0=48.4\,$mm. 
The diagonal distance between the surfaces of the rods is $2r_0=19\,$mm and the vertical 
distance between the surface of the rods is $8.5\,$mm, to allow for access to the MOT cooling laser beams.
The electrode radius is $r_e = 8.8$ mm, giving a ratio $r_e/r_0 = 0.9$, slightly smaller than the ideal ratio of 1.1468 according to Ref.~\cite{Denison:1971}. 

%Designing a Paul trap for use in a hybrid trap with a MOT requires a 
%careful balance between competing factors. The Paul trap must have good 
%optical access for the MOT trapping beams, since the capture velocity 
%is proportional to the size of the beams. This requirement suggests 
%that the Paul trap should be as large as possible, but if the trap 
%is too large, the proximity of the grounded vacuum chamber walls 
%interferes with the rf trapping potentials. 
%In addition, there is an optimal ratio of the electrode radius $r_e$ to 
%the diagonal distance from the center of the trap to the rods, $r_0$, 
%to minimize the contributions 
%of higher-order multipoles in the radial trapping potentials, which is 
%$r_e/r_0=1.1468$ \cite{Denison:1971}. In our trap, $r_e=8.8\,$mm and the ratio $r_e/r_0 = 0.9$.   
Our trap also deviates from the ideal Paul trapping potentials because the central 
segment is longer than the optimal length for a quadratic axial confining potential. 
Instead, the static axial confinement is a superposition of a quadratic and a quartic 
potential. As shown in our experiments and by the
  results of our simulations
  reported in Appendix B, trapping is possible
  even for electrode geometries that result in
  trapping potentials that have a strong
  quartic component and thus deviate considerably
  from the ideal quadrupole trapping potential.
  The price, as shown in Appendix B, is the
  emergence of deterministic chaos, even on the
  single-ion level, which reduces, and
  ultimately defines, the trapping volume of the LPT.

The number of ions in the Paul trap can be destructively measured using a 
channel electron multiplier (CEM). By putting DC voltages on the endcaps in a 
dipole configuration, the ions can be directed out of the trap along the axial 
direction toward the CEM. The programmed CEM extraction sequence
captures a constant fraction of the number of ions trapped
immediately before extraction.
This extraction efficiency is built into our calibration \cite{Goodman:2015},
which thus allows us to determine the number of ions
immediately before extraction.
We use a custom LabVIEW virtual instrument (VI) to control the experimental 
timing and to record the measurements. 

The basic experimental scheme is to allow the MOT to load to 
steady state from the background vapor in the presence of the photoionizing (PI) beam, then the
ion trapping potentials are turned on. This way the loading rate from the MOT is
constant during the entire time the ion trap is being loaded.
The ions are extracted after a set delay time of $0.001$ s after the loading ends
and the ion number is measured by the CEM during a $1$ s interrogation time. 
Since we can only measure the ion number destructively, we take 
many measurements at different loading times to create a time 
series of the ion number as a function of loading time. 
 
\begin{figure}
\centering
\includegraphics[width=\columnwidth]{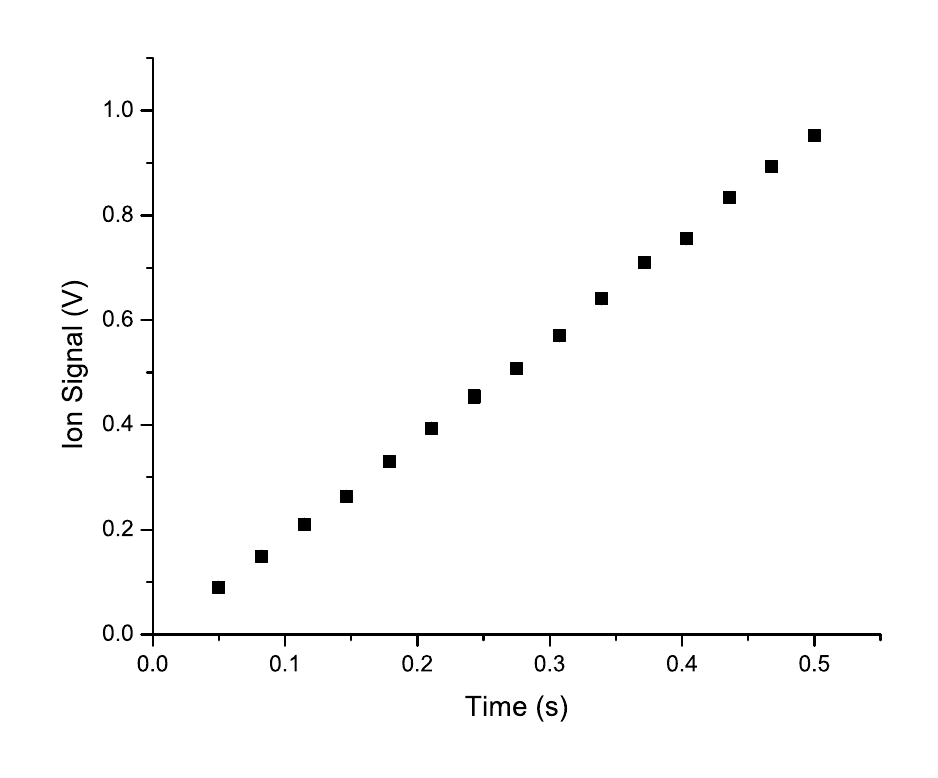}
\caption{\label{fig:loadrate} (Color online) 
Ion signal as a function of time for 
short loading times, which minimizes the effects from losses. 
The ion signal is proportional to the number of trapped ions; 
the slope is the loading rate from the MOT. Where the error bars (showing statistical errors) are not seen, they are smaller than the corresponding plot symbols. 
       }
\end{figure}

It is impossible to completely remove the ion loss mechanisms of the LPT, 
so the loading rate can never be measured independently of the loss rate. 
Whatever the exact form of the differential equation for the loading of the LPT, 
the loss rate depends on the number of trapped ions and the loading rate 
is independent of the number of trapped ions. Therefore, to measure the loading rate, 
the ion number is measured for very short loading times 
($\approx 0.05\,$s or shorter),  
where the loading can be described by [see (\ref{eq:tradmod})] 
\begin{equation}
\label{approxload}
\frac{dN}{d\tau}\approx\Lambda.
\end{equation}
An example of this is shown in 
Fig.~\ref{fig:loadrate}. We see that in the limit of small loading times
the number of ions trapped is a linear function of loading time.
  
%
%-----------------------------------------------------------------------
%--------------------------------------------------------------------------
%
  
To measure the steady-state ion number, the loading time is set 
to be long enough to ensure that the ion population in the trap has 
reached equilibrium. At the lowest loading rates this time could be up to 3 minutes; 
at the highest loading rates the trap is saturated in milliseconds.
Several consecutive measurements of the ion number are taken at a 
single loading time, then averaged to find the steady-state ion number. 
The loading rate is controlled by changing the $405\,$nm PI beam intensity 
and the process is repeated until the accessible part of 
the steady-state 
ion number as a function of loading rate is mapped out. With no PI 
beam present, no ion signal is measured. 
%Since we are directly measuring the ions in the trap 
%and the theory is agnostic to the actual loading mechanism 
%so long as the ions come from the MOT \cite{Blumel:2015}, 
%this shouldn't affect the agreement with theory.
%For the trap settings we used 
%during this study, no detectable number 
%of Na$_2^{+}$ ions was trapped.

The CEM measures a voltage that is proportional to the ion number, 
but the exact proportionality depends on the high-voltage gain applied to it. 
The most straightforward way to calibrate the CEM voltage would be to use an  
optical method to count the ion population and record the corresponding 
CEM voltage. However, because there are no optically accessible transitions 
in Na$^{+}$, the number of ions could not be measured directly. Furthermore, our CEM 
is designed to have a large bias current, which allows it to detect large ion signals. 
Consequently, the CEM can only operate in an analog 
detection mode and cannot be calibrated using pulse counting methods 
like those used in Ref.~\cite{RaviAPB:2012}. 

Instead, we used the same indirect method that we have used previously \cite{Goodman:2015}, 
which we will briefly describe here. The ion loading rate is measured two 
ways and then compared. First, the method described in Fig.~\ref{fig:loadrate} is used, with the 
assumption that the fraction of ions measured by the CEM,
 whatever it may be, does not change. The second method 
compares the loading of the MOT as a function of time
with and without the PI beam present. The increase of the loss rate in the 
presence of the $405\,$nm beam is equivalent 
to the loading rate of 
the Paul trap, under the assumption that every ion created 
from the MOT is trapped. This assumption is good when the MOT is smaller than 
the trapping volume of the Paul trap. To convert the PMT voltage to units 
of number-of-atoms, we use the two-level atom model to determine the 
excited-state population of the type-II MOT. One would expect this 
approximation to be especially poor in the case of the type-II MOT, 
where multiple hyperfine levels 
in the excited state play a role in the pumping transition. 
We find however that the two-level model used with a modified 
saturation intensity of $37.6\,$mW/cm$^2$, 
compared to the theoretical saturation 
intensity of $13.4\,$mW/cm$^2$, 
fits quite well. 
 
%
%-----------------------------------------------------------------------
\begin{figure}
\centering
\includegraphics[width=\columnwidth]{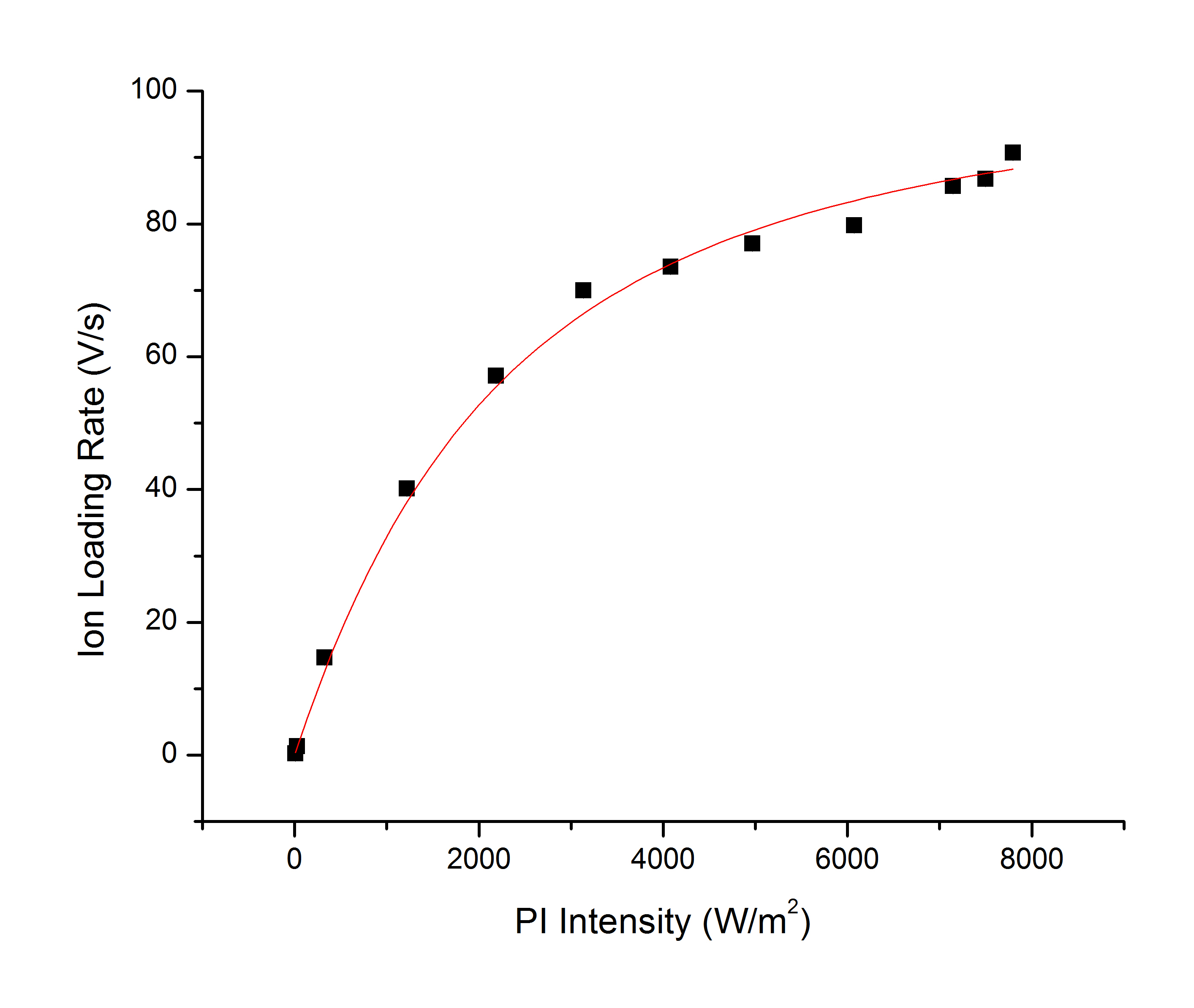}
\caption{\label{fig:cal} (Color online) 
Two-way measurement of the ion loading rate 
as a function of the intensity of 
the photo-ionizing laser for the purpose of 
finding the calibration factor between the ion signal and the number of ions. Measurements using the CEM (black squares) are compared to measurements of the atom loss rate from the MOT in the presence of the photoionization laser (solid red line) multiplied by a fitted scaling factor, which is the reciprocal of the CEM calibration factor. Where the error bars (showing statistical errors) are not seen, they are smaller than the corresponding plot symbols.  
       }
\end{figure}
%--------------------------------------------------------------------------
%

The two ion-loading rates are plotted and compared using a one-parameter fit, 
where the fitting parameter is the number of volts per ion. As seen in 
Fig.~\ref{fig:cal},
the fit shows good agreement with the data for all CEM bias voltages. 
This justifies the assumptions required for the calibration. 

\section{Experimental Results}
\label{EXP}

As discussed in Sec.~\ref{APP}, 
the electric 
potential in our LPT is not a pure, ideal 
quadrupole potential, but has a significant 
admixture of a quartic component. As discussed 
further in Appendix B, 
the quartic component in our potential leads to 
single-ion chaos, which renders our trap unstable 
in the radial direction from about $r=5\,$mm on, 
where $r=\sqrt{x^2+y^2}$ is the radial 
distance from the axis of the LPT. 
This might suggest that a proper description of 
our LPT is possible only in terms of a sum of 
quadratic and quartic terms. However, since 
up to $r=5\,$mm the quartic terms are small 
compared to the quadratic component, a simpler 
description of our LPT potential as a quadratic 
potential with a cut-off at 
$r=\hat R_{\rm cut}\approx 5\,$mm is possible. 
In this approximation, for $r<\hat R_{\rm cut}$, 
the electric potential of our LPT, 
written in SI units, is a pure 
quadrupole potential, given by 
\cite{Sivarajah:2012,Goodman:2012}: 
\begin{equation}
\phi(\vec r, \tau) = V_{\rm rf}\cos(\Omega \tau) 
\left( \frac{x^2-y^2}{r_0^2}\right) + 
\frac{\eta V_{\rm end}}{z_0^2} 
(z^2-\frac{1}{2}x^2 - \frac{1}{2}y^2), 
\label{LPTLS1}
\end{equation}
where $\vec r = (x,y,z)$ is the position vector of an ion in the 
trap, $\tau$ is the time, $V_{\rm rf}$ is the rf 
voltage applied to the electrodes of the trap, 
$\Omega = 2\pi f$ is the angular frequency of the applied 
rf voltage ($f$ is the lab frequency), $r_0$ and 
$z_0$ are defined in Sec.~\ref{APP}, 
$\eta=0.3$ is a dimensionless efficiency parameter, 
and $V_{\rm end}$ is the voltage applied to the end-segments of
the trap.  
For $r >\hat R_{\rm cut}$ the form of 
the potential is not needed, since ions are 
rapidly ejected from the trap (see Appendix B) as soon as they 
cross the chaos border at $r=\hat R_{\rm cut}$. 
Measured with respect to $\hat R_{\rm cut}$, 
and in pseudo-potential approximation 
(see Appendix A), the depth of the LPT 
potential is given by 
\begin{equation}
D = \left[ 
\frac{e^2 V_{\rm rf}^2}{m\Omega^2 r_0^4} - 
\frac{e\eta V_{\rm end}}{2 z_0^2} 
\right]\, \hat R_{\rm cut}^2.
\label{DEPTH}
\end{equation}
Experimentally, the trap depth $D$ can
be changed by changing the rf amplitude $V_{\rm rf}$,
the angular rf frequency $\Omega$ (via its dependence 
on the lab frequency $f=\Omega/(2\pi)$), 
or the end-cap potential $V_{\rm end}$. 
In the experiments reported in this paper, 
$V_{\rm end}=30\,$V 
is kept constant and only 
$V_{\rm rf}$ and $f$ are varied. 
Equation (\ref{DEPTH}) was derived for a single
trapped ion; if a second ion is trapped, 
the Coulomb repulsion changes the effective trap
depth each ion experiences. 
The anti-trapping space-charge effect from multiple trapped
ions makes the trap depth no longer 
expressible analytically. However, 
as supported by our analysis of ion numbers in 
the saturation region II below, the functional dependence of
the trap depth on $V_{\rm rf}$ and $f$ 
appears to be the same.
  
It is convenient to express the trap depth in terms of the 
single-particle stability parameter 
\begin{equation}
q = \frac{4eV_{\rm rf}}{m\Omega^2 r_0^2}
 \label{eq:q}
\end{equation}
so that
\begin{equation}
D = \left[ 
\frac{q e V_{\rm rf}}{4 r_0^2} - 
\frac{e\eta V_{\rm end}}{2 z_0^2} 
\right]\, \hat R_{\rm cut}^2.
\label{DEPTH-1}
\end{equation}
We also define the dimensionless loading rate
\begin{equation}
\lambda = 2\pi \Lambda / \Omega, 
\label{dimlam}
\end{equation}
i.e., the number of ions loaded per rf cycle.

\begin{figure}
\centering
\includegraphics[width=\columnwidth]{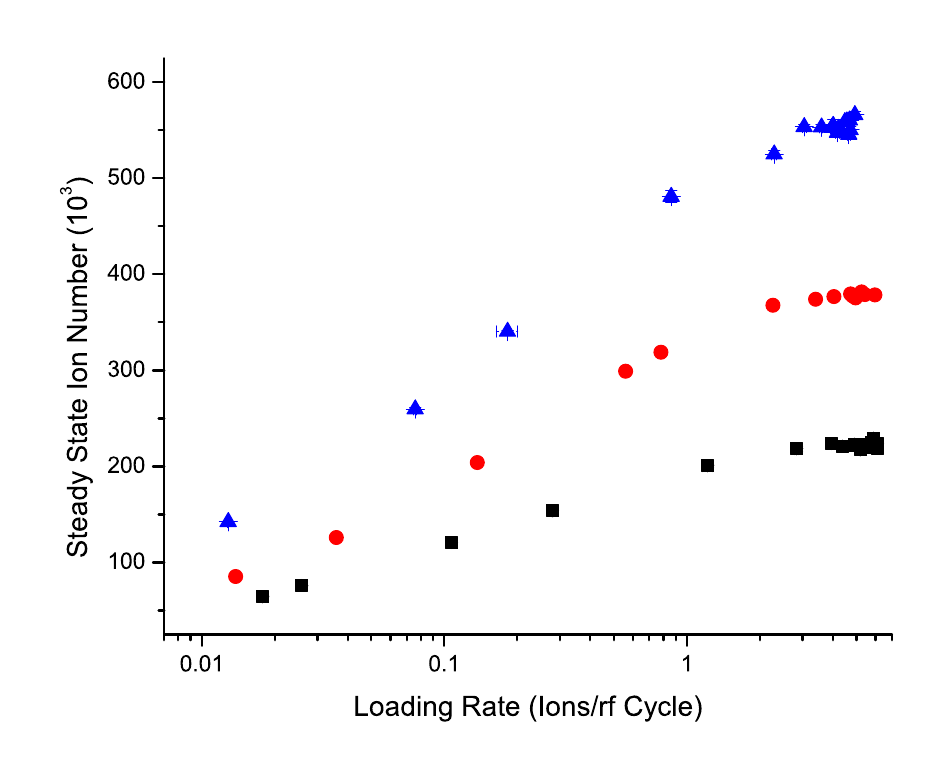}
\caption{\label{fig:conq}(Color online) Loading curves taken at constant $q=0.3$ and plotted on a lin-log scale. The trap settings were $V_\mathrm{rf}=13$ V and $f=450$ kHz for the squares (black), $V_\mathrm{rf}=16$ V and $f=500$ kHz for the circles (red), and $V_\mathrm{rf}=19.5$ V and $f=550$ kHz for the triangles (blue). Where the error bars (showing statistical errors) are not seen, they are smaller than the corresponding plot symbols.}
\end{figure}
 
In Fig.~\ref{fig:conq}, loading curves of the steady-state ion 
number as a function of loading rate $\lambda$
are shown for three sets of rf parameters that each result 
in $q=0.3$. In each case, both the
monotonic rise in region I and the plateau of 
region II are visible, predicted and previously
observed in [21]. 
Since $q$ is kept constant in 
all three cases shown in Fig.~\ref{fig:conq}, the depth of 
the LPT potential in these three cases is most conveniently 
evaluated according to (\ref{DEPTH-1}). In addition, 
(\ref{DEPTH-1}) shows that, for constant $q$, 
$D$ is independent of the frequency 
and depends only on $V_{\rm rf}$ and geometric constants 
of the LPT. 
Because of the quadratic form (\ref{LPTLS1}) 
of the LPT potential, 
the trapped ion cloud is an ellipsoid 
with semi-major axes equal to $\hat R_{\rm cut}$ in the 
$x$ and $y$ directions. In the $z$ direction we 
have $D=m\omega_z^2 \hat Z_{\rm cut}^2$, where 
$\hat Z_{\rm cut}$ is the extent of the ion cloud 
in the $z$ direction and $\omega_z$ 
is the pseudo-oscillator frequency in 
the $z$ direction. Since $\omega_z$ is determined 
by the static potential due to the end-caps of 
the LPT, $\omega_z$ is a constant and, therefore, 
$\hat Z_{\rm cut}\sim \sqrt{D}$. Thus, the 
volume of the trapped ion cloud is 
$V=(4\pi/3)\hat R_{\rm cut}^2\hat Z_{\rm cut}\sim 
\hat R_{\rm cut}^2 \sqrt{D}$. According to 
Poisson's equation of electrostatics, the 
density $\rho$ of the trapped ions is proportional 
to the Laplacian of the trapping potential, i.e., 
according to (\ref{LPTLS1}), 
$\rho\sim\nabla^2 \phi\sim V_{\rm rf}^2/\Omega^2$. 
This is all we need to predict, up to 
a proportionality constant, the number, $N$, of 
stored particles in the LPT. On the basis 
of the above discussion we have 
\begin{equation}
N=\rho V   \sim 
\hat R_{\rm cut}^2 V_{\rm rf}^2 D^{1/2}/\Omega^2 . 
\label{I-Number}
\end{equation}
More details on the derivation of (\ref{I-Number}) 
can be found in Sec.~\ref{5B}. 
 
We can now use (\ref{I-Number}) for a 
consistency check of our experimental results 
in Fig.~\ref{fig:conq}. 
Denoting 
by $N_{13}$, $N_{16}$, and $N_{19.5}$, the particle 
numbers corresponding to the cases $V_{\rm rf}=13\,$V, 
$V_{\rm rf}=16\,$V, and $V_{\rm rf}=19.5\,$V,
respectively,
in their asymptotic regimes, we may take their 
ratios $N_{13} : N_{16} : N_{19.5}$, which we can 
predict on the basis of (\ref{I-Number}), even 
without knowledge of the proportionality constant 
in (\ref{I-Number}). 
Indeed, assuming that $\hat R_{\rm cut}$ depends 
only weakly on $V_{\rm rf}$ and $\Omega$, 
the cut-off radius $\hat R_{\rm cut}$ cancels
when taking ion-number ratios 
on the basis of (\ref{I-Number}). 
Thus, the computation of ratios involves only 
$V_{\rm rf}$, $\Omega$, 
and known geometric constants. 
This way we obtain 
the explicit, analytical prediction 
$N_{13} : N_{16} : N_{19.5} = 1:1.63:2.48$. 
This prediction may be compared with the actual 
ion numbers in the plateau regimes read off 
from Fig.~\ref{fig:conq}. With 
$N_{13}\approx 225,000$, 
$N_{16}\approx 380,000$, and 
$N_{19.5}\approx 550,000$, 
we obtain 
$N_{13} : N_{16} : N_{19.5} = 1:1.69:2.44$. 
This is in excellent agreement with the 
theoretical prediction. As mentioned above, 
taking ratios has 
the advantage of eliminating $\hat R_{\rm cut}$, 
which is difficult to obtain directly in our experiments, 
since the trapped Na$^+$ ions are dark. 
 
\begin{figure}
\centering
\includegraphics[width=\columnwidth]{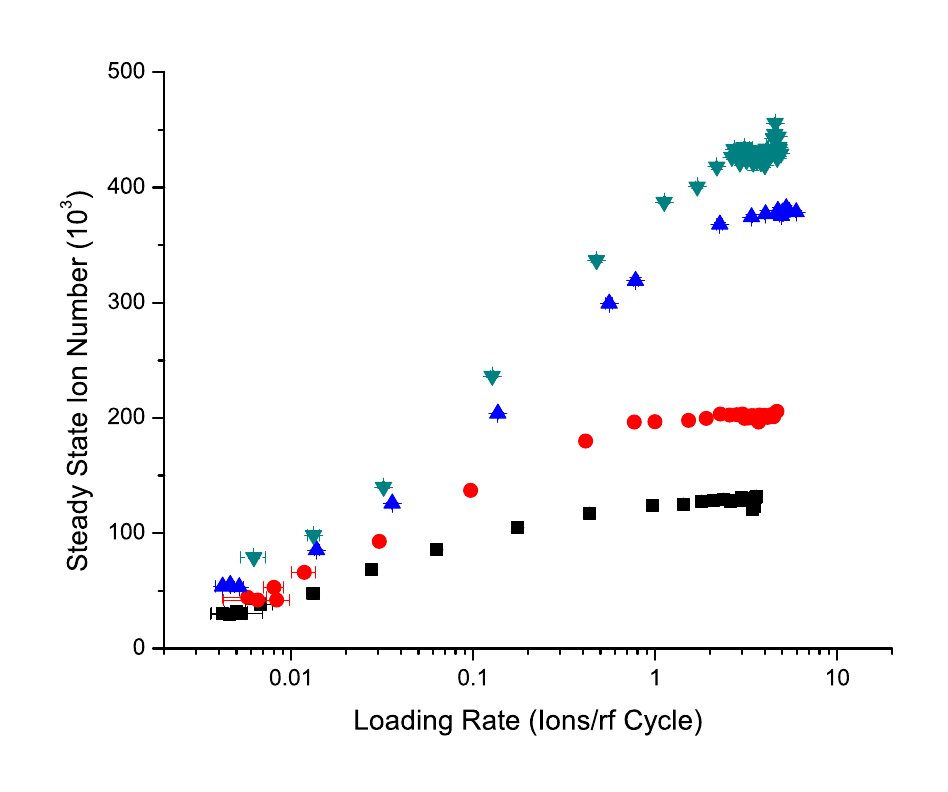}
\caption{\label{fig:conV}(Color online) Loading curves taken at constant $V_\mathrm{rf}= 16$ V and plotted on a lin-log scale. The trap settings were $q=0.22$ and $f=580$ kHz for the squares (black), $q=0.26$ and $f=535$ kHz for the circles (red), $q=0.30$ and $f=500$ kHz for the upright triangles (blue), and $q=0.37$ and $f=450$ kHz for the inverted triangles (turquoise). The plots are in order of their trap depths. Where the error bars (showing statistical errors) are not seen, they are smaller than the corresponding plot symbols.}
\end{figure} 
 
Region I and region II can likewise be seen in the curves 
shown in Fig.~\ref{fig:conV}, which are taken
at a constant rf amplitude, $V_{\rm rf} = 16\,$V. 
Shown in Fig.~\ref{fig:conV} are four loading curves with 
$f=580\,$kHz, $535\,$kHz, $500\,$kHz, and $450\,$kHz. 
Using the frequencies as labels, we read off 
$N_{580}\approx 130,000$, 
$N_{535}\approx 200,000$, 
$N_{500}\approx 380,000$, and 
$N_{450}\approx 430,000$, 
which yield the ratios 
$N_{580} : N_{535} : N_{500} : N_{450} = 
1:1.54:2.92:3.31$. The theoretical prediction 
for these ratios, according to (\ref{I-Number}), is 
$N_{580} : N_{535} : N_{500} : N_{450} =
1:1.57:2.14:3.29$. Just like for the 
results shown in Fig.~\ref{fig:conq}, and except for the curve with 
$f=500\,$kHz, the experimental ratios match the 
theoretical predictions very well.  
At present it is not clear why the third curve, at 
$f=500\,$kHz, is an outlier in this sequence. 
This is the more puzzling that this curve is the 
same as the corresponding curve shown in Fig.~\ref{fig:conq}, 
where it fits the sequence in Fig.~\ref{fig:conq} 
very well. That this 
curve does not fit well in Fig.~\ref{fig:conV} 
is also immediately obvious from 
the visual context in Fig.~\ref{fig:conV}. This curve 
produces a gap in region II of Fig.~\ref{fig:conV}, whereas 
a more even spacing was expected, mirroring the 
small decrements in frequencies corresponding to 
the four curves shown in Fig.~\ref{fig:conV}. 
 
\begin{figure}
\centering
\includegraphics[width=\columnwidth]{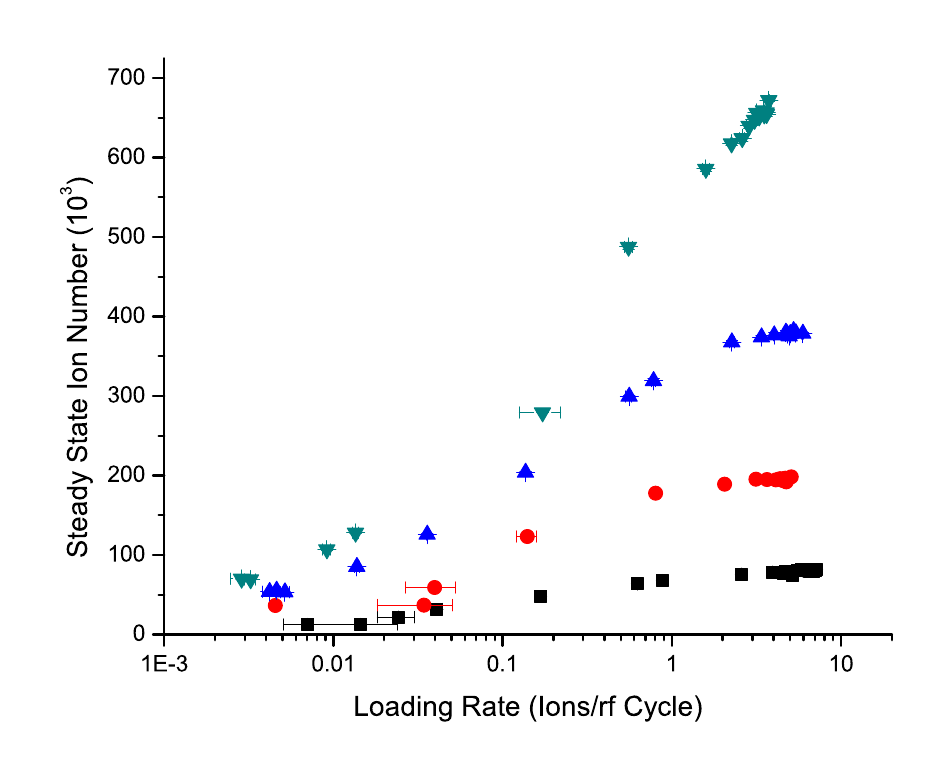}
\caption{\label{fig:conf}(Color online) Loading curves taken at constant $f = 500$ kHz and plotted on a lin-log scale. The trap settings were $q=0.22$ and $V_\mathrm{rf}=12$ V for the squares (black), $q=0.26$ and $V_\mathrm{rf}=14$ V for the circles (red), $q=0.30$ and $V_\mathrm{rf}=16$ V for the upright triangles (blue), and $q=0.37$ and $V_\mathrm{rf}=20$ V for the inverted triangles (turquoise). The plots are in order of their trap depths. Where the error bars (showing statistical errors) are not seen, they are smaller than the corresponding plot symbols.}
\end{figure} 
 
Several loading curves were also taken at a constant rf frequency; 
they are shown in
Fig.~\ref{fig:conf}. Akin to Figs.~\ref{fig:conq} and \ref{fig:conV}, 
the three lower curves all show both 
region I and region II. Only the loading curve taken at 
the largest rf voltage ($V_{\rm rf}=20\,$V) does not look 
like it has reached saturation (region II) yet, an 
impression confirmed by our ratio test to be conducted 
next. In the case of Fig.~\ref{fig:conf} there is a wrinkle in 
our theoretical analysis of the case $V_{\rm rf}=12\,$V in that for the experimental 
parameters the  density (\ref{DEPTH-1}) comes out 
negative, which means that we do not obtain 
a real square root in (\ref{I-Number}) and therefore 
$N$ cannot be predicted. This is not a disaster. 
It simply means that for $V_{\rm rf}=12\,$V and 
$f=500\,$kHz, our trap is operated so closely to 
the global instability border of the LPT that 
the pseudo-potential analysis (see Appendix A) 
is not accurate enough in this borderline case to make accurate  
predictions. For our ratio test we opted to ignore this 
case and normalize to the second curve in 
Fig.~\ref{fig:conf}, i.e., the case $V_{\rm rf}=14\,$V, 
$f=500\,$kHz, for which we obtain a positive 
trap depth of substantial magnitude 
for which our pseudo-potential analysis is valid. 
Using voltages as labels, like we did in 
the case of Fig.~\ref{fig:conq}, 
we predict 
$N_{14} : N_{16} : N_{20} =
1:1.95:4.65$.
From Fig.~\ref{fig:conf} 
we read off 
$N_{14}\approx 190,000$, 
$N_{16}\approx 380,000$, and 
$N_{20}\approx 660,000$, which results 
in the experimental ratios 
$N_{14} : N_{16} : N_{20} = 
1:2.00:3.47$. Similarly to the cases 
discussed in connection with Figs.~\ref{fig:conq} and \ref{fig:conV}, the 
ion-number ratio of the two cases of 
Fig.~\ref{fig:conf}, which are confidently 
in region II, is very 
close to its predicted value. In contrast, 
the predicted ratio for $N_{20}/N_{14}$ is much larger 
than the experimentally observed ratio, confirming our suspicion that 
for $V_{\rm rf}=20\,$V and at $\lambda=3\,$ions/rf-cycle 
the loading curve in this case is still 
climbing (still in region I), and its saturated 
ion number, expected to occur 
at higher values of $\lambda$ than those 
shown in Fig.~\ref{fig:conf}, 
will eventually be larger than $N=660,000$. 
 
The question arises whether we may simplify 
the expression 
(\ref{DEPTH}) [(\ref{DEPTH-1}), respectively] 
for the density $D$, 
perhaps by neglecting the term proportional 
to $V_{\rm end}$, in order to turn (\ref{I-Number}) 
into a more concise formula. 
Alas, as the case $V_{\rm rf}=12\,$V in 
Fig.~\ref{fig:conf} vividly illustrates, this is not possible. 
The two terms, i.e., the terms involving 
$V_{\rm rf}$ and $V_{\rm end}$ in 
(\ref{DEPTH}) [(\ref{DEPTH-1}), respectively], 
are of similar magnitude, which 
makes it impossible to neglect one with respect to the other. 
Thus, the square-root behavior in (\ref{I-Number}) 
is essential. 

Our theoretical estimates and predictions above are based on 
a single-ion picture (the pseudo-potential analysis 
presented in Appendix A), and do not include any 
space-charge- or many-body effects. 
The overall excellent agreement of our predictions 
for the ion-number ratios in region II of the 
cases shown in Figs.~\ref{fig:conq}, \ref{fig:conV}, and \ref{fig:conf} leads us 
to conclude that in our LPT experiments these effects 
are either negligible or lead to a simple renormalization 
of our expression for $N$ that results in an overall 
constant that cancels upon taking ratios. 
In Sec.~\ref{5B}, based on the single-particle 
pseudo-potential picture, we make predictions of 
the absolute magnitude of $N$ in region II, 
which agree very well with the experimentally 
observed values. This might argue for the 
space-charge- and many-body effects to be negligible. 
However, since $\hat R_{\rm cut}$ enters these 
formulas multiplicatively, we cannot be sure 
whether the renormalization constant is not 
simply absorbed in our effective $\hat R_{\rm cut}$, 
used in Sec.~\ref{5B}. Only direct 
experimental observation of $\hat R_{\rm cut}$ 
can resolve this issue. This, however, 
due to the optical darkness 
of the Na$^+$ ions used in our experiments, 
is currently beyond our experimental capabilities. 
Nevertheless, the excellent agreement of 
the experimental ion-number ratios in region II 
with our theoretical predictions supports 
the validity of our experimental 
LPT ion-loading curves. 

\section{Simulations}
\label{SIM}

Model simulations 
have already been done \cite{Blumel:2015} that confirm the existence 
of the four different dynamical regimes. In addition it was shown in \cite{Blumel:2015} that 
the phenomenon is robust with respect to the statistical distribution of time between loading events, 
temperature, loading mechanisms, and 
the geometry of the absorbing boundary. So here the emphasis is not so much on 
proving the existence of the four dynamical regions, or 
their robustness, but to see whether 
the simulations can qualitatively, and to some extent quantitatively, describe the 
experimentally observed characteristics of regions I and II.

 Since independence of the loading statistics has already been 
demonstrated in \cite{Blumel:2015}, we focus in this paper on the case of 
uniform loading statistics. Qualitatively, our results stay valid, 
which we checked explicitly, if different loading statistics, 
such as Gaussian statistics, are used. 

The Newtonian equations of motion of a 
singly charged ion of mass $m$ in the trap is 
\begin{equation}
m \frac{d^2\vec r}{d\tau^2} = - e \vec\nabla \phi(\vec r,\tau), 
\label{LPTLS2}
\end{equation}
which, written out in components, results in 
\begin{equation}
m\frac{d^2}{d\tau^2} \left( 
\begin{matrix} x \cr y \cr z \cr \end{matrix} \right)
= \left( 
\begin{matrix} -2 V_{\rm rf}e\cos(\Omega\tau)\frac{x}{r_0^2} + 
               \frac{\eta e V_{\rm end}}{z_0^2} x \cr 
               2 V_{\rm rf}e\cos(\Omega\tau)\frac{y}{r_0^2} + 
               \frac{\eta e V_{\rm end}}{z_0^2} y \cr
               -\frac{2\eta e V_{\rm end}}{z_0^2} z \cr 
\end{matrix}  \right) . 
\label{LPTLS3}
\end{equation}
Defining the dimensionless time 
\begin{equation}
   t = \left( \frac{\Omega}{2} \right) \tau , 
\label{LPTLS4}
\end{equation}
the set of equations may be written as 
\begin{equation}
 \left( 
\begin{matrix} \ddot x \cr \ddot y \cr \ddot z \cr \end{matrix} \right)
= \left( 
\begin{matrix} -2q\cos(2t)x + bx  \cr 
                2q\cos(2t)y + by  \cr
               -2b z \cr 
\end{matrix}  \right) ,  
\label{LPTLS5}
\end{equation}
where the dots indicate differentiation with respect 
to dimensionless time $t$, the dimensionless control parameter $q$ is defined in (\ref{eq:q}),
and 
\begin{equation}
b = \frac{4e\eta V_{\rm end}}{m\Omega^2 z_0^2} . 
\label{LPTLS7}
\end{equation}
If more than one ion of charge $e$ and 
mass $m$ are stored in the trap, the ions 
interact via the Coulomb force, resulting in the 
following set of coupled equations: 
\begin{equation}
\left(   \begin{matrix}   
\ddot x_i + 2q\cos(2t) x_i - b x_i \cr
\ddot y_i - 2q\cos(2t) y_i - b y_i \cr
\ddot z_i + 2bz_i  \cr 
          \end{matrix} \right) 
   = \sum_{\substack{j=1 \\ j\neq i}}^{N} 
   \frac{   \vec r_i - \vec r_j} {|\vec r_i - \vec r_j|^3 }, 
\label{LPTLS8}
\end{equation}
where $i=1,\ldots,N$ counts the number of particles 
in the trap at time $t$, and 
$\vec r$ is measured in units of 
\begin{equation}
l_0 = \left( \frac{e^2}{\pi\epsilon_0 m \Omega^2}\right)^{1/3},  
\label{LPTLS9}
\end{equation}
where $\epsilon_0$ is the electric permittivity of the vacuum. 

It is the introduction of the unit of length (\ref{LPTLS9}) 
that allows us to normalize the coefficient in 
front of the Coulomb force on the right-hand side 
of (\ref{LPTLS8}) to 1, and thus arrive at the set of equations 
(\ref{LPTLS8})
that depends only on the two dimensionless parameters 
$q$ and $b$. $q$ is an adjustable control parameter that, 
depending on the trap voltage $V_\mathrm{rf}$ and frequency $f$, 
may be set to a value in the interval $0 < q \lesssim 0.9$, 
where $q\approx 0.9$ is the Mathieu 
instability limit \cite{Abramowitz:1964}. It is shown in Appendix A 
that for given $q$, in order
to achieve trapping, the parameter $b$ has to satisfy
$0 < b < q^2/2$.  
 
Numerically, because of the repulsive Coulomb interactions
 in (\ref{LPTLS8}), and even
 for a large 
number $N$ of trapped ions, 
the coupled set of equations (\ref{LPTLS8}) is 
well conditioned. Therefore, a standard 
4th order Runge-Kutta method \cite{Press:1992} 
is enough to reliably 
integrate (\ref{LPTLS8}). 
 
While our numerical model (\ref{LPTLS8}) 
captures the essence of 
the experimental LPT, and its parameters 
$q$ and $b$ are adjusted 
to their experimental values, our model 
is nevertheless an idealization. The electrodes 
in our experiment are not hyperbolic surfaces, as 
required if (\ref{LPTLS8}) is expected to be exact, and 
while (\ref{LPTLS8}) assumes a quadratic potential in 
$z$ direction, the potential in our LPT also has a
quartic component \cite{Goodman:2012,Sivarajah:2012}. Still, the proportions 
of the numerical trap, as expressed in (\ref{LPTLS8}), 
are correct and we expect that our model captures 
the essential parts of the physics in our experimental LPT. 

A final comment concerns the number of particles we are able to 
simulate compared with the number of particles in our experimental trap. 
In order to accumulate sufficient statistics, and given 
our computer resources, we found that 2,000 simultaneously stored 
ions are a practical upper limit for our numerical simulations. 
Although orders of magnitude smaller than 
the experimental number of particles in our trap, 2,000 particles 
is not a small number, and using scaling relations, to be 
discussed below, we are able to compare our simulations not 
just qualitatively, but also quantitatively with our experimental 
results.  
 
Our simulations proceed in the following way. 
For a given loading rate $\lambda$ we generate a time 
sequence $t_j, j=1,...,M$ of loading events, which 
have a Poissonian distribution whose average corresponds 
to the specified loading rate $\lambda$. For each 
individual parameter setting we check that $M$ 
is large enough so that we are deeply in the saturated 
regime where we are able to extract the steady-state
number of ions, $N_s$, with excellent 
statistics. Denoting by $\langle \ldots \rangle_t$ 
the time average in the saturated regime, 
we also compute the statistical spread 
$\Delta N_s=\sqrt{\langle N_s^2\rangle_t-
\langle N_s\rangle_t^2}$, which characterizes the ion-number 
fluctuations in the saturated regime due to 
the Poissonian loading process. 
At each loading event $t_j$ a new ion with zero 
initial velocity is created in the trap 
at a random location inside of a spherical loading zone 
of radius $\hat R_{\rm load}$ (in SI units), 
representing the creation of ions from the MOT via the 
photoionizing $405\,$nm laser. Between loading events, 
i.e. for $t$ in 
$t_j < t < t_{j+1}$, we integrate the ion trajectories 
in the trap, including the newly created ion, according 
to the system (\ref{LPTLS8}). Once $t_{j+1}$ is reached, 
we eliminate all ions from the trap whose positions 
at $t_{j+1}$ lie beyond a 
pre-specified absorbing boundary B. 
In \cite{Blumel:2015} we already showed that the qualitative 
shape of the $N_s(\lambda)$ curves does not 
depend on the geometry of B. Therefore, 
making use of this freedom, 
we chose in this paper a cylindrical absorbing 
boundary B with radius $\hat R_{\rm cut}$ in the 
$x$-$y$ direction and total length $2z_0$ in the $z$ direction, i.e., ions are
 absorbed in the $z$ direction if they exceed
 $|z|=\hat Z_{\rm cut}=z_0=24.2\,$mm.
 
In addition 
to $q$, all we need for 
our simulations are the parameters $b$, 
$R_{\rm load}=\hat R_{\rm load}/l_0$,  
$R_{\rm cut}=\hat R_{\rm cut}/l_0$, and
 $Z_{\rm cut}=\hat Z_{\rm cut}/l_0$.
The radius of the loading zone $\hat R_{\rm load}$ 
is defined by the size of the type-II MOT, 
which, according to Ref.~\cite{Goodman:2015}, is 
$r_a=0.75\,$mm. 
The electrodes are positioned at 
$r_0=9.5\,$mm, which is the upper bound for $\hat R_\mathrm{cut}$. 
According to the discussion above, we cannot 
simulate the full-sized experimental LPT, since it typically holds 
more ions than we are able to realistically simulate. 
Accordingly, we simulate a scaled-down version 
of our LPT in which all linear dimensions 
are scaled by a factor $0<\sigma < 1$, resulting in 

\begin{align}
R_{\rm load} &= \sigma \hat R_{\rm load}/l_0 
             = 88.4\, \sigma\,  [f({\rm MHz})]^{2/3} , \cr
R_{\rm cut} &= \sigma \hat R_{\rm cut}/l_0 
             = 1119.1\, \sigma\, [f({\rm MHz})]^{2/3}, \cr
Z_{\rm cut} & = \sigma \hat Z_{\rm cut}/l_0
             = 2827.2\, \sigma\, [f({\rm MHz})]^{2/3}. 
\label{LPTLS11}
\end{align}

As discussed in Sec.~\ref{EXP}, in a harmonic trap, i.e., a trap with time-independent
quadratic trapping potentials in
all three directions, and at zero temperature, 
the charge density $\rho_{\rm el}$ in the trap is 
constant, which follows immediately from 
$\rho_{\rm el} \sim  \nabla^2 \phi$. In this case 
the particle number in the trap 
scales with the volume of the trap, i.e., 
\begin{equation}
N \sim \sigma^3.
\label{Nscale-1}
\end{equation}
Although our experimental trap is not 
exactly harmonic, and the temperature is finite, 
we nevertheless expect that (\ref{Nscale-1}) 
holds to a good approximation. 
Therefore, in order to compare with our 
experimental results, we scale 
$N_s$ obtained from our 
simulations according to 
\begin{equation}
N_s^{\rm scaled} = N_s^{\rm simulated} / \sigma^3, 
\label{Nscale-2}
\end{equation}
which allows a direct comparison between 
the results of our 
simulations with experimental results of 
the saturated number of ions in the trap. 

\begin{figure}
\centering
\includegraphics[width=\columnwidth]{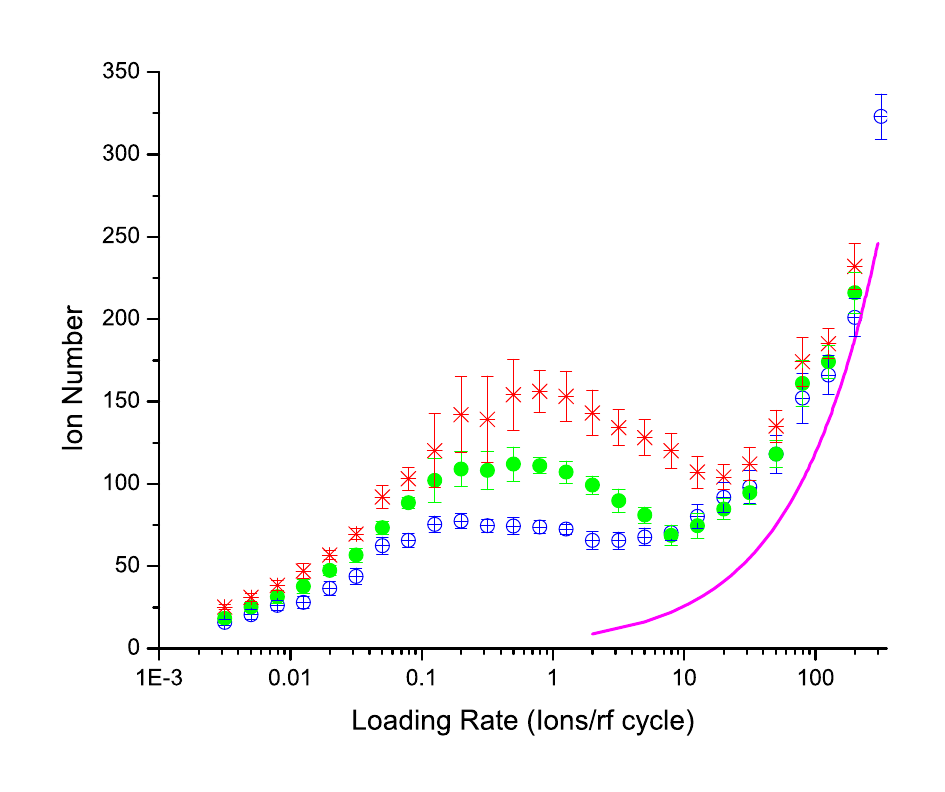}
\caption{\label{figG} (Color online) 
Simulation results for the model LPT performed at 
constant $q=0.3$ for three different combinations 
of rf voltage and frequencies. 
$f=450\,$kHz, $V=13\,$V: open, blue circles; 
$f=500\,$kHz, $V=16\,$V: filled, green circles;
$f=550\,$kHz, $V=19.5\,$V: red asterisks. 
Scale parameter: $\sigma=1/40$. 
These simulation 
results may be compared with 
the three corresponding 
LPT experiments shown in Fig.~\ref{fig:conq}. The heavy solid line (purple) is the analytical 
result for $N_s(\lambda)$ in region IV \cite{Blumel:2015}. The lengths of the error bars, equal to
 $2\Delta N_s$, characterize the statistical
 fluctuations of the ion number in the saturated
 regime.
       }
\end{figure}

We are now ready for the numerical simulations. Since they are expensive, 
we focused on simulating
 the constant-$q$ 
experiments, performed with $q=0.3$, 
and described in Sec.~\ref{EXP}. 
Figure~\ref{figG} shows the results of our simulations. 

%
%-----------------------------------------------------------------------
%--------------------------------------------------------------------------
%

\begin{figure}
\centering
\includegraphics[width=\columnwidth]{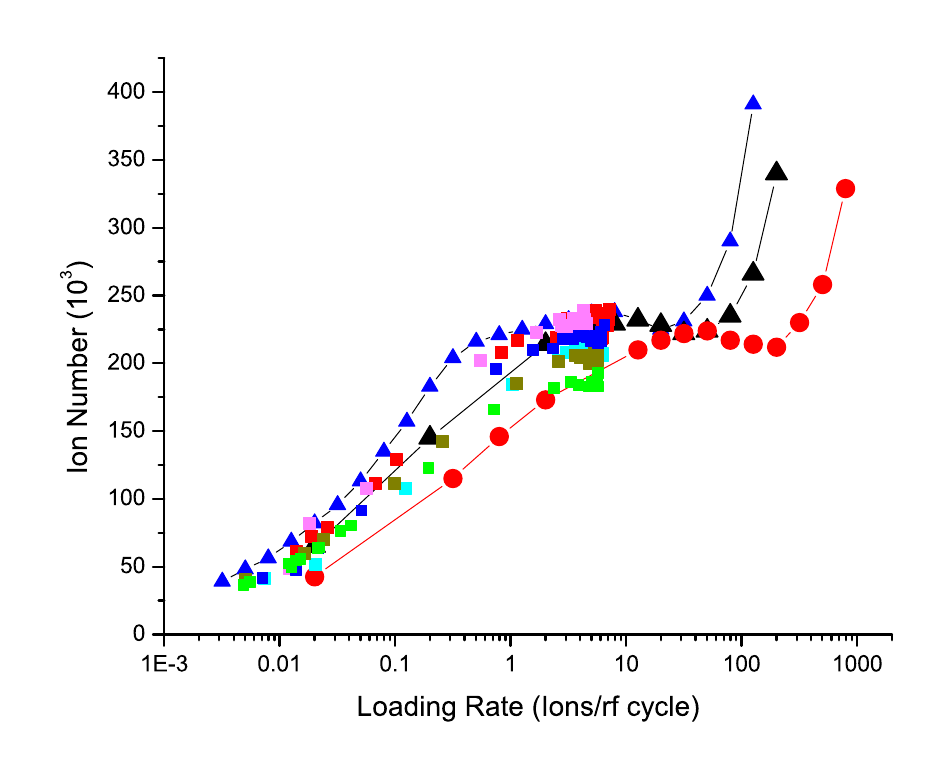}
\caption{\label{fig:simex}(Color online) Simulations of loading curves at three scaling factors: $\sigma = 1/10$ small triangles (blue), $\sigma = 1/8$ large triangles (black), and $\sigma = 1/5$ circles (red). Also shown are a number of experimental loading curves (squares) at $q=0.30$, $V_\mathrm{rf}=13$ V, and $f = 450$ kHz. The scatter of the experimental data points gives
an idea of the statistical and systematic variations
in our experimental loading data. For these simulations, $\hat R_{\rm cut}=4.5\,$mm was chosen,
which, as a result of many simulations, akin to those shown
in this figure, turned out to yield the best agreement
with the experimental results. Where the experimental error bars (showing statistical errors) are not seen, they are smaller than the corresponding plot symbols.
}
\end{figure}

Compared with the experimental results shown in Fig.~\ref{fig:conq}
we see that in our simulations the maximum of region II occurs 
at a loading rate which is about a factor 10 lower than 
in the experiments. However, the location of the region-II
maximum depends on the scaling factor $\sigma$ and shifts
to higher loading rates as $\sigma$ approaches 1 where the
simulated ion trap size becomes identical to the experimental
Paul trap. This effect can be seen in Fig.~\ref{fig:simex}, where
simulations are plotted at several different values of $\sigma$, along
with several experimental curves. Additionally, the simulations in 
Fig.~\ref{fig:simex} have been scaled up in ion number [see (\ref{Nscale-2})]. The 
simulated ion curves match the experiments in the ion number, 
but the regions occur at different ion loading rates. The reason for 
this discrepancy is not clear. It is possible that effects not included
in the idealizations made for the simulations, such as stray fields causing
excess micromotion or the presence of the grounded vacuum chamber,  
are the cause. 

\section{Analytical Theory}
\label{THEORY}

In this section we present an analytical 
theory for regions I and II. In Sec. V.A 
we show that in region I the saturated 
ion number $N_s(\lambda)$ follows a 
power law in $\lambda$, which we confirm 
experimentally. We also compute the approximate 
exponent of the power law, which agrees well 
with our experiments. In Sec. V.B we 
present a theory for region II. This theory 
explains the plateau behavior of $N_s(\lambda)$ 
in region II and also fits the temporal behavior 
of $N_s(\lambda,t)$ better than all other 
theories so far described in the literature. 

\subsection{Region I}

In this subsection we present a simple, analytically 
solvable model for the steady-state 
ion population $N_s(\lambda)$ as a function of loading rate $\lambda$. 
Our model predicts monotonic 
power law behavior in region I. Under certain reasonable assumptions, based 
on heating rates obtained from 
molecular-dynamics simulations of non-neutral plasmas published 
in the literature \cite{Tarnas:2013}, 
the exponent derived from our 
analytical model is close to the exponent observed in our experiments. 
Since our analytical calculations 
assume spherical symmetry, our analytical 
results for region I are primarily applicable to 
the three-dimensional quadrupole Paul trap (3DPT). 
Comparing with experiment, we found that these results 
also hold well for the LPT. 
 
In region I we need to consider the stationary
state in which for each particle loaded 
one particle escapes. 
In the stationary state the spatial probability distribution,  
$\rho(\vec r\,)$, 
of the ions 
in the trap is approximately Gaussian with a width that is proportional 
to $\sqrt{T}$, where $T$ is the temperature. Since, according to $E\sim kT$, 
where $k$ is the Boltzmann constant, 
the average energy $E$
of a stored ion is proportional to $T$, the width of the spatial Gaussian 
is proportional to $\sqrt{E}$. 
 
In order to obtain analytical results in closed form, we replace the Gaussian 
distribution by a flat distribution with a sharp cut-off, i.e. we represent 
the spatial density $\rho(\vec r\,)$ by 
a homogeneous sphere according to 
\begin{equation}
\rho(\vec r\,) = 
\begin{cases}    \displaystyle\frac{3N_s}{4\pi w^3},    &  |\vec r\,| \leq w,   \\ 
                0,                         &  |\vec r\,| >  w, 
\end{cases} 
\label{rho}
\end{equation}
where $N_s$ is the number of ions in steady state, 
\begin{equation}
w = \alpha\sqrt{E} 
\label{weq}
\end{equation}
is the radial width of the density distribution, and 
$\alpha$ is a constant. Since a Gaussian is a steeply 
descending function in the wings, the approximation (\ref{rho}) 
is benign and does not change 
the scaling of $N_s(\lambda)$ in $\lambda$.  
 
\begin{figure}
\centering
\includegraphics[width = \columnwidth]{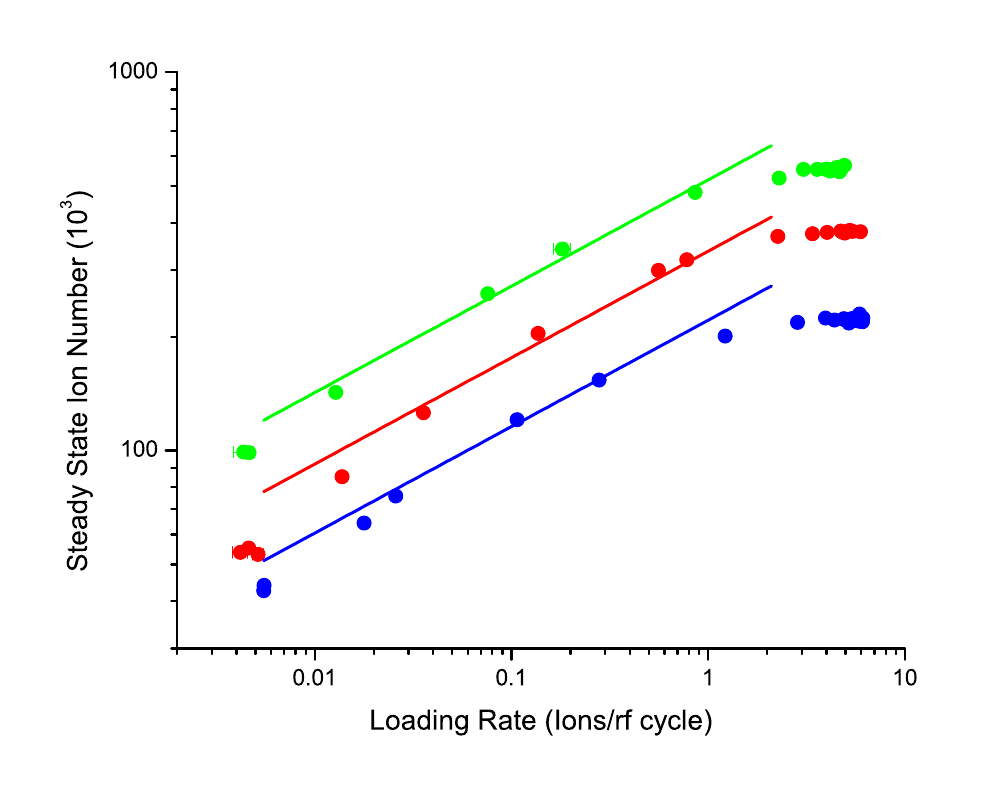}
\caption{\label{fig:R1Fits}(Color online) The data points from Fig.~\ref{fig:conq} shown with
fit using $N_s=A\lambda^{0.281}$. We see that the power-law form of the fit function, as predicted
by our analytical model, fits the data in region I very well. Where the error bars (showing statistical errors) are not seen, they are smaller than the corresponding plot symbols.
}
\end{figure}

We start with the situation in which the probability sphere, due to 
rf heating \cite{Tarnas:2013,Nam:2014,Blumel:1988}, has expanded beyond the location  $R_{\rm cut}$ of the absorbing boundary B, just far enough for the total excess probability 
beyond $R_{\rm cut}$ to integrate to 1 particle. Denote by $\kappa$ the 
heating rate per particle. Then, the energy $E$ per particle is 
\begin{equation}
E = E_0 + \kappa \Delta \tau,
\label{Eequ}
\end{equation}
where $\Delta \tau$ is the time that has passed since the last loading 
(ion creation) event and $E_0$ is the energy per particle 
immediately after the last loading event. Since we are in the steady state, 
exactly $\Delta \tau=1/\Lambda$ has passed on average 
between ion creation and ion loss, 
where $\Lambda$ is the loading rate. 
Therefore, from (\ref{Eequ}), 
\begin{equation}
E = E_0 + \frac{\kappa}{\Lambda} , 
\label{EDt}
\end{equation}
and, according to (\ref{weq}), 
\begin{equation}
w = \alpha 
\left(E_0+ \frac{\kappa}{\Lambda}\right)^{1/2}.
\label{wDt}
\end{equation}
The excess width is 
\begin{equation}
\delta w = w - R_{\rm cut} = \alpha \left(E_0+\frac{\kappa}{\Lambda} 
\right)^{1/2} - R_{\rm cut}. 
\label{dw}
\end{equation}
In steady state, the excess width $\delta w$ corresponds to 
exactly 1 particle. Therefore, denoting by $\delta V$ 
the volume of the shell 
of width $\delta w$: 
\begin{align}
1 &= \rho\delta V = \rho 4\pi R_{\rm cut}^2 \delta w \nonumber \\
&= 
\left( \frac{3N_s}{4\pi R_{\rm cut}^3}\right)
4\pi R_{\rm cut}^2 \left[ 
\alpha \left(E_0+\frac{\kappa}{\Lambda}\right)^{1/2} - R_{\rm cut}\right] 
\nonumber 
  \\
&= \frac{3N_s}{R_{\rm cut}} \left[ 
\alpha \left(E_0+\frac{\kappa}{\Lambda}\right)^{1/2} - R_{\rm cut}\right] . 
\label{exc}
\end{align}
In region I, i.e., 
for small $\Lambda$, the dominant term in (\ref{exc}) is $\kappa/\Lambda$. 
Therefore, in region I, we may approximately write: 
\begin{equation} 
1 = \frac{3N_s\alpha}{R_{\rm cut}}\left(\frac{\kappa}{\Lambda}\right)^{1/2}. 
\label{approx}
\end{equation}
In order to compute the dependence of 
$N_s$ on $\Lambda$, we need to know how the 
heating rate $\kappa$ depends on $N_s$. 
In order to answer this question we turn 
to Fig.~2 of \cite{Tarnas:2013}. This 
figure shows the heating rate $H$ of 
$N$-ion clouds in steady state 
as a function of cloud size $\hat s$. 
This figure is relevant since, 
up to the choice of units, 
$\kappa$ and $H$ are identical. 
Therefore, 
the $N$-scaling of $\kappa$ is the same as the 
$N$-scaling of $H$. 
Although Fig.~2 of Ref.~\cite{Tarnas:2013} 
was computed for a 3DPT, it 
nevertheless gives us a first idea on the 
$N_s$ scaling of $\kappa$ for the LPT 
that we focus on in this paper. 
Since our trap has an effective radius $R_{\rm cut}$, 
we need to extract heating rates $H$ from 
Fig.~2 of \cite{Tarnas:2013} as a function of $N$ 
at constant cloud size $\hat s$. 
The most striking feature of the heating 
data shown in 
Fig.~2 of \cite{Tarnas:2013} is that the heating rate 
curves for different $N$ are parallel to each other 
and have about the same spacing when doubling 
the number of particles $N$. Therefore, 
since Fig.~2 of \cite{Tarnas:2013} shows $H$ on 
a log scale, both features combined show that, 
at given $\hat s$, independently of $\hat s$, 
$H$ follows a power law in $N$. On the basis 
of the data displayed in Fig.~2 of \cite{Tarnas:2013}, we 
find, at constant $\hat s$: 
\begin{equation}
H \sim N^{10/3}. 
\label{hrate}
\end{equation}
Therefore, because of $\kappa\sim H$, we obtain: 
\begin{equation}
\kappa = \beta N_s^{10/3},
\label{kaprate}
\end{equation}
where $\beta$ is a constant. 
Using this result in (\ref{approx}), 
we obtain 
\begin{equation}
1 = \frac{3\alpha \beta^{1/2} N_s^{8/3}}{R_{\rm cut} \Lambda^{1/2} } .
\label{nearfin}
\end{equation}
Using (\ref{dimlam}),
we solve (\ref{nearfin}) for $N_s$ 
in terms of $\lambda$: 
\begin{equation}
N_s = \left[\frac{R_{\rm cut}}{3\alpha (2\pi\beta/\Omega)^{1/2}}\right]^{3/8}\ 
\lambda^{3/16}. 
\label{fin}
\end{equation}
Thus, this simple model predicts $N_s\sim \lambda^{0.188}$, which may be 
compared with the experimental 
region-I result $N_s\sim \lambda^{0.281}$. Since the numerical value of the exponent predicted
by our model
depends on the scaling of the
heating rate $\kappa$ in $N$,
which was not separately determined for our
LPT, the value of the exponent predicted by (\ref{fin}) is
less important than the prediction that
$N_s$ follows a power law. Therefore,
we treat the value of the exponent
as a fit parameter. Fits
 of the data from Fig.~\ref{fig:conq}
using this model can be seen in Fig~\ref{fig:R1Fits}. The difference
in the value of the exponent is likely due to the difference in heating rate between
the ideal 3DPT of \cite{Tarnas:2013} and the heating rate in the 
experimental LPT.

\begin{figure}
\centering
\includegraphics[width=\columnwidth]{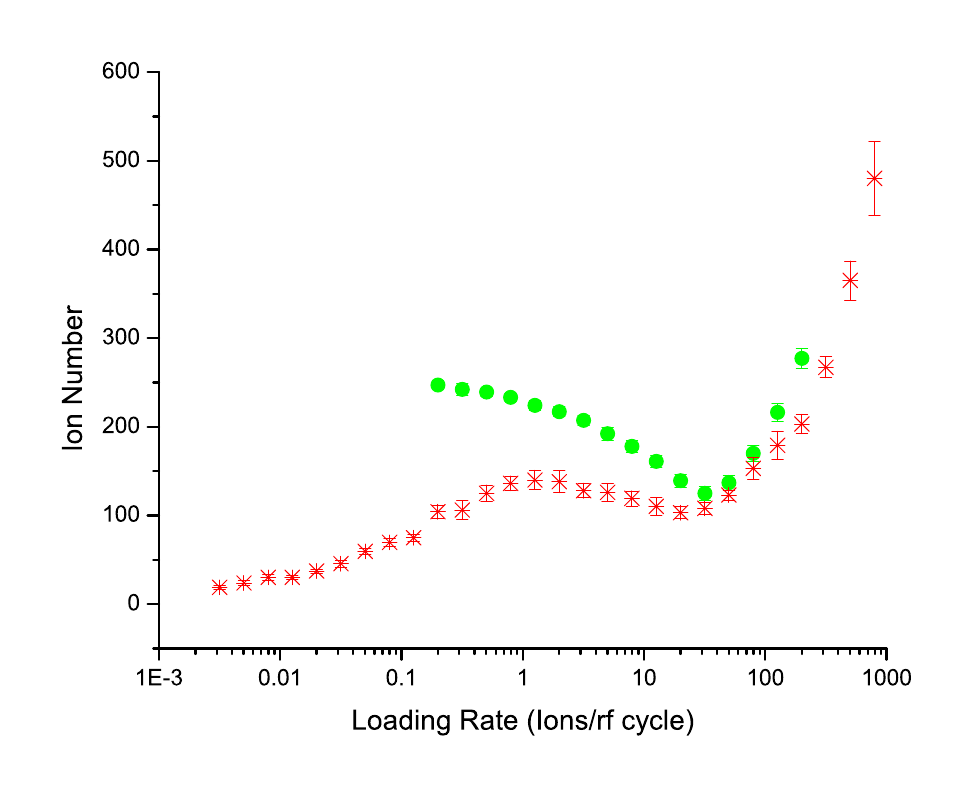}
\caption{\label{fig:mm} (Color online) Simulation of the LPT:
Comparison between the time-dependent model with
rf switched on (asterisks)
and the pseudo-potential model with only
the pseudo-potential present (filled circles).
We simulated the case $V_{\rm rf}=16\,$V,
$f=450\,$kHz, with scaling factor
$\sigma=1/40$. The lengths of the error bars, equal to
 $2\Delta N_s$, characterize the statistical
 fluctuations of the ion number in the saturated
 regime. Where error bars are not seen, they are
 smaller than the plot symbols.
       }
\end{figure}

Since rf heating is one of the 
central ingredients in this model, both the prediction of a power law 
in itself and the approximate agreement of the power law exponent with 
our experimental results indicate that rf heating is the factor that 
governs the behavior of $N_s(\lambda)$ in region I. 
This is corroborated by Fig.~\ref{fig:mm}, which shows a
comparison between $N_s$ obtained as a result of
solving the fully time-dependent set of equations
of motion (\ref{LPTLS8}), i.e., the equations
of motion including rf heating (asterisks
in Fig.~\ref{fig:mm}) and $N_s$ obtained as a result
of solving the time-independent pseudo-oscillator
equations of motion (\ref{APP-PP14}), which, due to the lack
of explicit time-dependence, are not capable of
simulating rf heating (filled circles
in Fig.~\ref{fig:mm}). Clearly, while the
pseudo-oscillator model is capable of
reproducing regions II, III, and IV, it
completely fails to reproduce
region I, which can only be attributed to a lack
of rf heating in the pseudo-oscillator model,
since otherwise this model contains all
many-body forces
exactly as in the full set of equations (\ref{LPTLS8}).
Thus, we proved conclusively that it is
rf heating that determines both the
power law and the power law's exponent
in region I.

\subsection{Region II}
\label{5B}

In region II, the steady-state ion number plateau is determined by 
the depth of the trap. The pseudopotential approximation, described in Appendix A, 
is valid in region II. The pseudopotential well is completely filled with ions in this region. 
The expression for the depth of the pseudopotential predicts a steady-state ion number
in region II of 
\begin{multline}
N_s^{II} = \frac{4\pi^3\epsilon_0m\times10^3}{3e^2}f[\mathrm{Mhz}]^2\times\\\hat R_{\rm cut}[\mathrm{mm}]^3q^2\sqrt{\frac{q^2}{4b}-\frac{1}{2}}.
\label{eq:R2N}
\end{multline}

There are no adjustable parameters in this expression. However,
 $\hat R_\mathrm{cut}$ is not easily determined for dark ions. The traditional
 derivations of the pseudopotential depth assume that $\hat R_\mathrm{cut} = r_0$
 and that the ions are only lost when they collide with or move beyond the
 trap electrodes \cite{Ghosh:1995,Major:2005}. This has been shown not to 
be the case in recent hybrid trap work \cite{Ravi:2012,Goodman:2015}. Finding
 $\hat R_\mathrm{cut}$ directly allows the trap depth
% \begin{equation}
% D_r = 1/2*m*(\omega_r)^2*(\hat R_mathrm{cut})^2
 %\label{eq:trapdepth}
 %\end{equation}
 to be calculated without making any approximations.  In principle, $\hat R_{\rm cut}$
  can be found in a variety of ways. 

The first is to follow the method used in \cite{Ravi:2012,Goodman:2015}, where the
idealized single-particle trap depth is equated to the energy of a simple harmonic oscillator
with spring constant $k=m\omega^2$, where $\omega$ is the secular frequency of 
the ionic motion. This method has the advantage of being simple, but the
drawback is that it relies on the single-particle trap depth, which is a credible
approximation, but may not be accurate enough. Indeed, this simple model
disagreed with subsequent hybrid trap experiments, described in \cite{Goodman:2015}, by 25\%.

A second method, mentioned in the caption of Fig.~\ref{fig:simex}, is to find the $\hat R_\mathrm{cut}$ 
value which makes the simulations best match the experimental results.  
This requires some computationally demanding simulations of the trapped
ions. 

A third method is to find an $\hat R_\mathrm{cut}$ that makes 
(\ref{eq:R2N}) fit best at one trap setting. For sodium ions, (\ref{eq:R2N}) becomes
\begin{equation}
N_s^{II} = 548344f[\mathrm{Mhz}]^2\hat R_\mathrm{cut}[\mathrm{mm}]^3q^2\sqrt{\frac{q^2}{4b}-\frac{1}{2}}.
\label{eq:R2Nb}
\end{equation} By selecting a single data set and finding the $\hat R_\mathrm{cut}$ 
that makes both sides of (\ref{eq:R2Nb}) approximately equal, 
we are able to fit nearly all of our data at least as 
well as the other methods, and in some cases much 
better, using a much simpler procedure (see Table~\ref{tb:R2}). 
The data set with the smallest ion number, which was taken at 
$q=0.22$, $V_\mathrm{rf}=16$ V, and $f=500$ kHz, is the one exception;
using these parameters in this model returns an imaginary number of ions. 
These settings also have the smallest number of ions at steady state in 
region II, i.e., approximately 80000, which is still a large number for 
most Paul trap experiments. It is also unclear what differentiates the
settings where this value of $\hat R_\mathrm{cut}$ fits well and the ones that do not. 

There has been no previous study of why $\hat R_\mathrm{cut}\neq r_0$, but it has
been hypothesized that it could be due to contributions from higher-order 
multipoles \cite{Alheit:1996}. In this paper we confirm that it is
  the admixture of the quartic component of the confining
  potential in $z$ direction that is responsible for
  the reduction of $\hat R_{\rm cut}$ to
  $\hat R_{\rm cut} < r_0$. In fact, while a single ion
  in an ideal trap, i.e., a trap in which only quadrupole
  potentials are present, is never chaotic, even when
  driven by the rf trap fields, a single ion
  in a potential with a quartic admixture shows a
  transition in space from regular,
  confined motion, to chaotic, unconfined motion (see Appendix B).
  This brings us to a fourth method for
  determining $\hat R_{\rm cut}$, described in detail
  in Appendix B. According to this method,
  $\hat R_{\rm cut}$ is identical with the
  single-ion chaos border. This means that
  for $r < \hat R_{\rm cut}$ the ion's motion
  is perfectly regular and the ion, in the absence
  of noise, is perfectly trapped. However,
  as soon as $r$ exceeds $\hat R_{\rm cut}$, the
  ion's motion becomes chaotic. As a consequence,
  the ion is free to explore spatial
  regions with $r > \hat R_{\rm cut}$, which
  quickly leads to an encounter with the electrodes
  at which the ion is absorbed. Thus,
  $\hat R_{\rm cut}$ is determined by a
  method that is based on the
  dynamics of a single ion and therefore
  allows a very quick and efficient
  determination of $\hat R_{\rm cut}$ for
  various trap settings. This method does not only
  have technical advantages for the determination
  of the value of $\hat R_{\rm cut}$. It also
  solves the puzzle of the very existence of
  $\hat R_{\rm cut}$, identifying its origin as
  a fundamental, purely dynamical effect, a chaos
  transition, whose exact location is determined
  by the strength of the admixture of higher
  multipole fields to the LPT's quadrupole trapping field.

\begin{table}
\caption{\label{tb:R2}The experimental ion number compared with the ion number predicted by (\ref{eq:R2Nb}) with $\hat R_\mathrm{cut}=3.74$ mm. The cells with N/A indicate an imaginary number predicted.}
\begin{ruledtabular}
\begin{tabular}{cccccc}
$q$ & f[MHz] & $b$ & \begin{tabular}{@{}c@{}}Predicted\\ Ion Number \end{tabular}& \begin{tabular}{@{}c@{}}Measured\\ Ion Number\end{tabular} & \begin{tabular}{@{}c@{}}Percent\\ Difference\end{tabular} \\
0.30 & 0.550 & 0.022 & 574273 & 566162 & 1.4 \\
0.30 & 0.500 & 0.026 & 387308 & 381897 & 1.4 \\
0.30 & 0.450 & 0.032 & 231851 & 229396 & 1.1 \\
0.22 & 0.500 & 0.026 & N/A       & 81654   & N/A \\
0.26 & 0.500 & 0.026 & 185254 & 198434 & 6.6 \\
0.37 & 0.500 & 0.026 & 882666 & 672169 & 31.3 \\
0.37 & 0.450 & 0.032 & 594951 & 455767 & 30.5 \\
0.26 & 0.535 & 0.023 & 271704 & 205591 & 32.2 \\
0.22 & 0.580 & 0.019 & 163397 & 132166 & 23.6
\end{tabular}
\end{ruledtabular}
\end{table}

The loading of the Paul trap as a function of time was also examined. 
In our previous work \cite{Blumel:2015}, the loading as a function of time in region IV
was determined from the equation
\begin{multline}
\label{eq:R4L}
t =-\frac{2\tilde{R}}{3\tilde{\lambda}^{1/3}}\left\{\ln{(\alpha-\tilde{N}^{1/2}})-\frac{1}{2}\ln{(\tilde{N}+\alpha\tilde{N}^{1/2}+\alpha^2)}\right. \\
\left. +\sqrt{3}\arctan{\left(\frac{2\tilde{N}^{1/2}+\alpha}{\alpha\sqrt{3}}\right)}-\sqrt{3}\arctan{\left(\frac{1}{\sqrt{3}}\right)}\right\},
\end{multline}
where $\alpha = \tilde{R}^{1/2}\tilde{\lambda}^{1/3}$, 
and the tilde indicates that the quantity corresponds
to the loading zone, which in this case is the MOT volume.
Therefore, for example, $\tilde{R}$ is the radius of the 
loading zone, which in this case is the MOT radius. 

\begin{figure}
\centering
\includegraphics[width=\columnwidth]{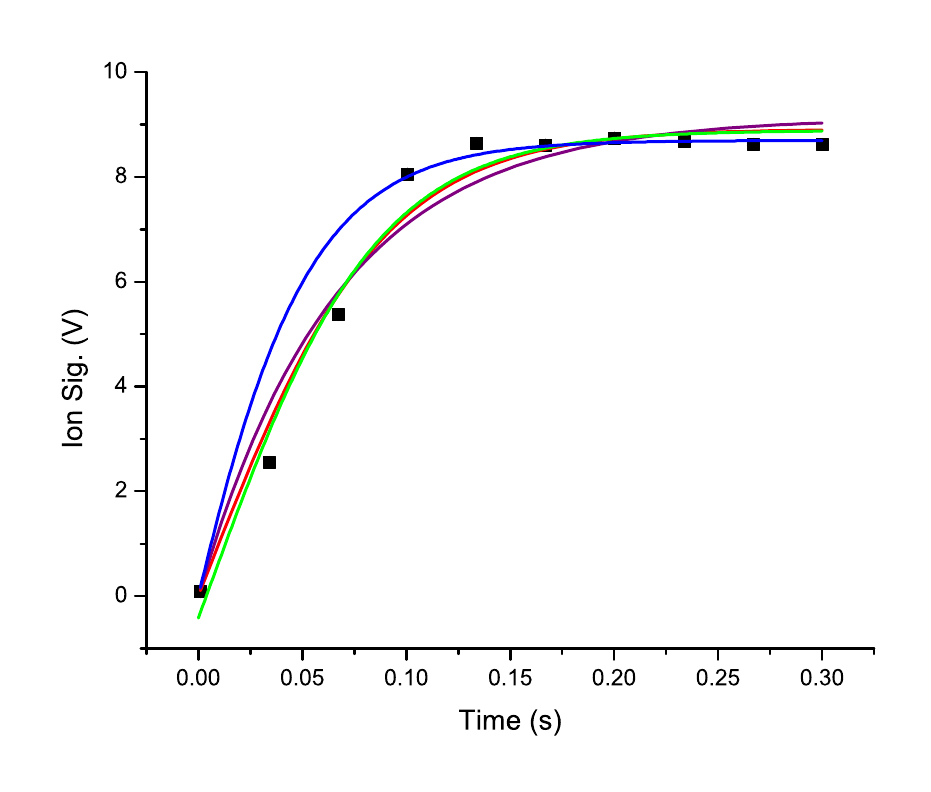}
\caption{\label{fig:NvT}(Color Online) Ion number vs. time fitted to the 
model (\ref{eq:R4L}) (blue), the traditional model (\ref{eq:tradmod}) (green), 
that model with $\ell_2 = 0$ (purple), 
and that model with $\ell_1 = 0$ (red). 
The traditional model displays the familiar 
overshoot-undershoot-overshoot pattern, 
but the model (\ref{eq:R4L})  fits both the knee and the plateau. Where the error bars (showing statistical errors) are not seen, they are smaller than the corresponding plot symbols.}
\end{figure}

Equation (\ref{eq:R4L}) was derived for region IV and one should not 
expect it to fit in regions I and II. Indeed, applying it to
plots at extremely small loading rates results in poor fits and the
prediction of imaginary ion numbers. However, in region II, (\ref{eq:R4L})
fits the loading data at least as well, if not better, 
than the traditional model (\ref{eq:tradmod}). This includes 
versions of the traditional model where $\ell_1=0$ or $\ell_2=0$, 
as can be seen in Fig.~\ref{fig:NvT}. All forms of the 
traditional model demonstrate the overshoot of the rise, undershoot of the knee, and
overshoot of the plateau seen also in Fig.~\ref{fig:tradfit}. While (\ref{eq:R4L}) overshoots the rise as well, it fits the knee and
the plateau better than the traditional model. Further, unlike the traditional model, it is derived from the
underlying physical process and depends on quantities that can be experimentally measured
independently of the model itself.

\section{Conclusion}
\label{CONC}

In this work, two of the four regions in the loading curve of a 
Paul trap, predicted in \cite{Blumel:2015}, have been confirmed
experimentally. Additionally, both simulations and analytical models
match reasonably well to the steady-state ion number and shape of the curves. The two
regions are relevant and accessible to the majority of Paul trap experiments.
 Also proposed are new, experimental and computational methods for finding 
 $\hat R_\mathrm{cut}$, a quantity necessary for finding the actual trap depth (a difficult experimental prospect), 
 the total collision rate constant for dark ions \cite{Lee:2013,Goodman:2015},
and for describing the behavior of the Paul-trap loading in region III \cite{Blumel:2015}. 
Further work remains to be done to fully understand the loading of
the Paul trap. An upcoming paper will examine the behavior of regions III and IV
through experiment, simulations, and analytical theory. 

\section{Acknowledgements}
W.W.S. would like to acknowledge NSF support (in part) from grant 1307874.  

\appendix 
\section*{Appendix A: LPT Pseudopotential}
\label{APP-PP} 
In this appendix we derive the single-particle pseudopotential for the 
LPT used in the analysis of our experiments. The pseudopotential is used 
in our numerical simulations to prove (a) that the power-law behavior
of the loading curves in region I is a consequence of
rf heating, which we prove in reverse by
demonstrating that the pseudo-potential equations
of motion, lacking an rf term,
cannot explain region I,
and (b) that the existence of region II
does not depend on the time-dependence of the rf 
drive of the trap, i.e., as demonstrated in Sec.\,\ref{5B}, 
the time-independent pseudopotential 
alone is capable of explaining the plateau in region II. 
 
We start with the equations of motion (\ref{LPTLS5}), 
where $x$, $y$, and $z$ are in units of $l_0$ [see (\ref{LPTLS9})], 
$t$ is in units of $2/\Omega$ [see (\ref{LPTLS4})], and 
$q$, 
$b$ are the dimensionless control 
parameters defined in (\ref{eq:q}) and (\ref{LPTLS7}), 
respectively.  
Focusing on the $x$ component of (\ref{LPTLS5}), and following 
the procedure outlined in \cite{Landau:2007}, we split the $x$ coordinate 
into a large-amplitude, slowly varying component $X(t)$, 
called the macromotion, and a small-amplitude, 
fast-oscillating component $\xi(t)$, called the 
micromotion according to 
\begin{equation}
x(t) = X(t) + \xi(t). 
\label{APP-PP1}
\end{equation}
Defining the cycle average 
\begin{equation}
\langle f(t) \rangle = \frac{1}{\pi} \int_{-\pi/2}^{\pi/2} f(t+t')\, dt', 
\label{APP-PP2}
\end{equation}
and in line with the physical meanings of $X$ and $\xi$, we assume 
\begin{align}
\langle X(t)\rangle &= X(t), \ \ \ \langle \xi(t) \rangle = 0, 
\nonumber \\ 
\langle \ddot X(t)\rangle &= \ddot X(t), 
\ \ \ \langle \ddot \xi(t) \rangle = 0  . 
\label{APP-PP3}
\end{align}
Focusing first on the time-dependent (rf) part of (\ref{LPTLS5})  
and using the decomposition (\ref{APP-PP1}), 
we have 
\begin{equation}
\ddot X(t) + \ddot \xi(t) = -2q \cos(2t) [X(t)+\xi(t)] . 
\label{APP-PP4}
\end{equation}
Because $\ddot \xi(t)$ dominates the left-hand side of 
(\ref{APP-PP4}) and $X(t)$ dominates the right-hand side, we 
may write approximately 
\begin{equation}
\ddot \xi(t) = -2q \cos(2t) X(t). 
\label{APP-PP5}
\end{equation}
Since, according to (\ref{APP-PP3}), $X(t)$ is assumed 
to be constant 
over one rf cycle, we may integrate (\ref{APP-PP5}) immediately, 
resulting in 
\begin{equation}
\xi(t) = \frac{q}{2} \cos(2t) X(t),  
\label{APP-PP6}
\end{equation}
where we set the integration constants to zero. This 
is necessary for consistency, since these constants lead to 
non-oscillating, slow terms that are assumed to be contained 
in $X(t)$. To compute $X(t)$, we take the cycle 
average of (\ref{APP-PP4}). Assuming that $X(t)$ and 
$\xi(t)$ are uncorrelated, i.e., $\langle X(t)\xi(t)\rangle=0$, 
and $\langle \cos(2t) X(t)\rangle = 0$, 
we arrive at 
\begin{align}
\ddot X(t) &= -2q \langle \cos(2t)\xi(t)\rangle = -2q \langle 
\frac{q}{2} \cos^2(2t)X(t)\rangle 
\nonumber \\ 
&= -\frac{q^2}{2} X(t),  
\label{APP-PP7}
\end{align}
where we used $\langle\cos^2(2t)\rangle=1/2$. 
This equation of motion for $X(t)$ may be derived 
from the potential 
\begin{equation}
U_{\rm eff}(X) = \frac{q^2}{4} X^2  
\label{APP-PP8}
\end{equation}
via 
\begin{equation}
\ddot X(t) = -\frac{\partial U_{\rm eff}(X)}{\partial X}. 
\label{APP-PP9}
\end{equation}
To obtain $U_{\rm eff}(X)$ in SI units, we multiply 
(\ref{APP-PP8}) with the unit of energy 
\begin{equation}
E_0 = \frac{ml_0^2\Omega^2}{4}. 
\label{APP-PP9a}
\end{equation}
The force proportional to $b$ in (\ref{LPTLS5}) may be 
derived from the potential 
\begin{equation}
U_{\rm stat}(X) = -\frac{b}{2} X^2. 
\label{APP-PP10}
\end{equation}
Notice that this potential is deconfining. 
Combining (\ref{APP-PP8}) and (\ref{APP-PP10}) results in 
the total pseudopotential 
\begin{equation}
U_{\rm pp}(X) = U_{\rm eff}(X) + U_{\rm stat}(X) = 
\left( \frac{q^2}{4} - \frac{b}{2}\right) X^2 
\label{APP-PP11}
\end{equation}
acting on the macromotion coordinate $X$ of an ion in the LPT. 
Since the $y$ equation of (\ref{LPTLS5}) is formally identical 
with the $x$ equation, we obtain immediately: 
\begin{equation}
U_{\rm pp}(Y) = \left( \frac{q^2}{4} - \frac{b}{2}\right) Y^2, 
\label{APP-PP12}
\end{equation}
where $Y$ is the macromotion coordinate of a trapped ion in $y$ 
direction. 
To obtain 
the pseudopotential for the $z$ coordinate of a trapped ion, 
all we need to do is to set $q$ to zero and replace $b\rightarrow -2b$ in the 
above derivations to obtain 
\begin{equation}
U_{\rm pp}(Z) = b Z^2, 
\label{APP-PP13}
\end{equation}
where $Z$ is the macromotion coordinate of an ion in $z$ direction. Clearly, in order
to achieve trapping in the $x$ and $y$ directions,
we need the pseudo-oscillator potentials in
$x$ and $y$ directions to be confining, which
requires the coefficients in front of the $X^2$ and
$Y^2$ terms in (A12) and (A13) to
be positive, which, in turn, requires
$b<q^2/2$. In order to achieve trapping in the
$z$ direction, we need $b$ in (A14) to be positive.
Combining these two conditions, we obtain the condition
\begin{equation}
0 < b < q^2 / 2
\label{globstab}
\end{equation}
as the condition for global stability
of the LPT in pseudo-potential approximation. 

On the basis of the LPT pseudopotentials 
(\ref{APP-PP11}), (\ref{APP-PP12}), and (\ref{APP-PP13}), 
we now obtain the set of equations of motion of a trapped 
ion in pseudopotential approximation: 
\begin{equation}
\left(   \begin{matrix}   
\ddot x_i + \frac{q^2}{2} x_i - b x_i \cr
\ddot y_i + \frac{q^2}{2} y_i - b y_i \cr
\ddot z_i + 2bz_i  \cr 
          \end{matrix} \right) 
   = \sum_{\substack{j=1 \\ j\neq i}}^{N} 
   \frac{   \vec r_i - \vec r_j} {|\vec r_i - \vec r_j|^3 }.  
\label{APP-PP14}
\end{equation}
Notice the change of sign in the $y$-equation part of 
(\ref{APP-PP14}) with respect to (\ref{LPTLS8}), which is 
consistent, since the rf field, on average, produces a confining  
force in $y$ direction, which, in the pseudopotential 
equations (\ref{APP-PP14}), requires a ``$+$'' sign in 
front of the $\frac{q^2}{2} y_i$ term. 
 
 \appendix 
\section*{Appendix B: The dynamical origin of $\hat R_{\rm cut}$}
\label{APP:B}
 In this appendix we show that $\hat R_{\rm cut}$
has a purely dynamical origin. It is
explained as a chaos border due to the quartic
admixture in the $z$ potential
$P(z)$ of the trap. A fit of
$P(z)$ on the axis of the trap yields
\begin{multline}
P(z) =
%3.23130
3.231
\times 10^{-5}z^4 -
%1.9854
1.985
\times 10^{-6} z^3\\
\left.  +
%3.3911754
3.391
\times 10^{-3} z^2+
%4.825917
4.826
\times 10^{-4} z +
%0.2690146329
0.269,\right.
\label{AB-1}
\end{multline}
where
$z$ is in mm and $P(z)$ is in volts.
Assuming cylindrical symmetry and neglecting
the small terms asymmetric in $z$ proportional to $z$
and to $z^3$, we extend $P(z)$ into
the $x$ and $y$ directions, i.e.,
$P(z)\rightarrow P(x,y,z)$, by requiring
$\nabla^2 P(x,y,z)=0$. We obtain
\begin{multline}
P(x,y,z) = 0.269V + \frac{1.986\,{\rm V}}{z_0^2}
\left( z^2-\frac{1}{2}r^2\right)\\\left. +
        \frac{11.082\,{\rm V}}{z_0^4}
        \left(z^4 + \frac{3}{8} r^4
                        - 3z^2 r^2\right),\right.
\label{AB-2}
\end{multline}
where $z_0=24.2\,$mm.
The extension $P(z)\rightarrow P(x,y,z)$
is unique, once
cylindrical symmetry is assumed. We are aware
of the fact that
cylindrical symmetry can be true only
close to the LPT's axis, since closer to the
rods, we have a four-fold symmetry, which breaks
rotational invariance around the LPT's $z$ axis.
However, close to the LPT's axis, (\ref{AB-2}) is
an acceptable analytical approximation, which,
for $r\lesssim 5\,$mm, differs
from the experimental $P(x,y,z)$ by less than
30\%.
%
%-----------------------------------------------------------------------
\begin{figure}
\centering
\includegraphics[width=\columnwidth]{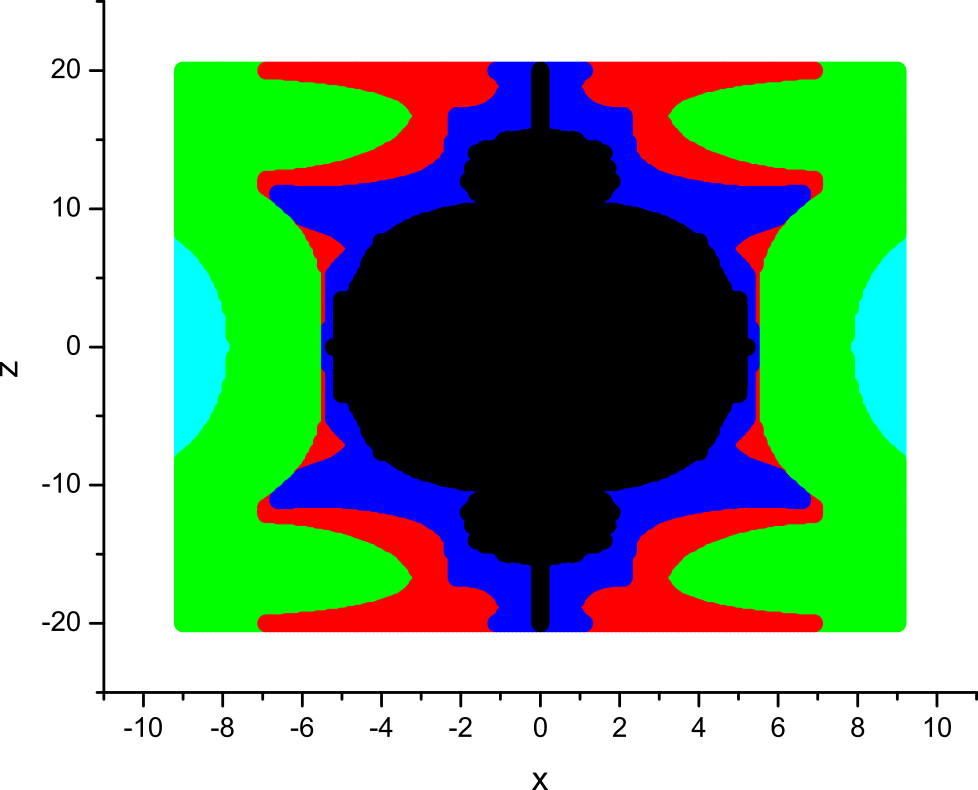}
\caption{\label{EFRACT} (Color online)
Color-coded fractal
of escape times. Denoting by $L$ the number of
rf cycles it takes for an initial condition
$(x,z)$ to reach $|x| \geq r_0=9.5\,$mm, the colors
code for $L<3$ (cyan),
$3 \leq L < 6$ (green),
$6 \leq L < 9$ (red),
$9 \leq L < 1,000$ (blue), and
$L > 1,000$ (black). The black area corresponds
to initial conditions that lead to ion trajectories
that never escape.
The black area is bounded in $x$ direction by
$\hat R_{\rm cut}\approx 5\,$mm. The points with
$x=0$, protruding from the fractal, are also shown
in black, since they
correspond to on-axis equilibrium points
that, too, never escape.
       }
\end{figure}
%--------------------------------------------------------------------------
%

On the basis of (\ref{AB-2}) we obtain the following
single-ion
equations of motion
\begin{equation}
\left(   \begin{matrix}
\ddot x + 2q\cos(2t) x - b_2 x + b_4\left[
 \frac{3}{2}(x^3+xy^2)-6xz^2\right] \cr
\ddot y - 2q\cos(2t) y - b_2 y + b_4\left[
 \frac{3}{2}(y^3+yx^2)-6yz^2\right] \cr
\ddot z + 2b_2 z + b_4 \left[
4z^3 - 6z(x^2+y^2) \right]  \cr
          \end{matrix} \right)
   = 0,
\label{AB-3}
\end{equation}
where
\begin{equation}
b_2 = \frac{1.986\,{\rm eV}}{m\pi^2 z_0^2 f^2}, \ \ \
b_4 = \frac{11.082\,{\rm eV}\,l_0^2}{m\pi^2 z_0^4 f^2}
\label{AB-5}
\end{equation}
and $l_0$ is defined in (\ref{LPTLS9}).
Integrating the system of equations (\ref{AB-3})
for many initial conditions, we found that the
single-ion dynamics governed by (\ref{AB-3}) exhibits
trapped and escaping trajectories. We illustrate this
in the following way. For $q=0.3$, $f=450\,$kHz (the case shown in
Fig.~\ref{fig:simex}), we determined the lifetimes $L$ (in rf cycles)
of 72,000 trajectories with initial conditions
$x_n=-9\,{\rm mm}+n\times 0.1\,{\rm mm}$, $n=1,\ldots,180$,
$y=0$, and
$z_m=-20\,{\rm mm}+m\times 0.1\,{\rm mm}$, $m=1,\ldots,400$.
The color-coded result is shown in Fig.~\ref{EFRACT}.
We see that the set of initial conditions that leads to
trajectories that never escape (black
region in Fig.~\ref{EFRACT}) has a finite area, extending
less than $5\,$mm in $x$ direction, while trajectories that
visit any of the colored regions quickly escape.
Therefore, from
Fig.~\ref{EFRACT}, we conclude that
$\hat R_{\rm cut}\approx 5\,$mm. This is consistent
with the value $\hat R_{\rm cut}\approx 4.5\,$mm
used in Fig.~\ref{fig:simex}.
By repeatedly zooming into the boundary of the black
region in Fig.~\ref{EFRACT} we checked explicitly
that the black region in Fig.~\ref{EFRACT}
has a fractal boundary
\cite{mandelbrot:1982},
which indicates that trajectories started close
to the boundary are transiently chaotic
\cite{lai:2011}.
Thus, $\hat R_{\rm cut}$ is identified as
a chaos border. Therefore, far from caused by any
non-controllable effects, such as
patch fields, stray fields, or noise
(although these effects certainly may modify
$\hat R_{\rm cut}$),
the reduced trapping capacity of our LPT,
characterized by $\hat R_{\rm cut}$, is
a purely deterministic, dynamic effect, which
is fundamentally related to the shape of
the trapping potential of our LPT in $z$ direction.
While the investigation of the properties of the
escape fractal shown in Fig.~\ref{EFRACT} is an
interesting project in itself, it is
beyond the scope of this paper and
not necessary
for the purpose of explaining the dynamical origin of
$\hat R_{\rm cut}$. We will report more results on
the escape fractal, including chaos and order in our LPT,
elsewhere.

%\bibliography{Bibliography}
%

\end{document}